\documentclass[a4paper,12pt]{article}
\usepackage{jheppub} 
\usepackage{tcilatex}
\usepackage{amsmath}
\usepackage{amsfonts}
\usepackage{amssymb}
\usepackage{multicol}
\usepackage{graphicx}
\usepackage{float}
\usepackage{caption}
\usepackage{xcolor}
\usepackage[utf8]{inputenc}

\title{Fluctuating Ensemble Averages and the BTZ Threshold}
\author[1,2]{R. Sammani,}
\author[1,2,3]{E.H Saidi,}
\author[1,2]{R. Ahl Laamara,}
\author[1,2,3]{L.B Drissi}

\affiliation[1]{ LPHE-MS, Science Faculty, Mohammed V University in Rabat, Morocco}
\affiliation[2]{Centre of Physics and Mathematics, CPM- Morocco}
\affiliation[3]{Hassan II Academy of Science and Technology, Kingdom of Morocco.}
\emailAdd{rajae$\_$sammani@um5.ac.ma}
\abstract{ Recent work shows fascinating links between ensemble averaging and the
Swampland program. In order to break the emerging global symmetries of the
ensemble averaging as dictated by the no global symmetries conjecture, one
may consider fluctuations away from the average given by deviations in the
Siegel-Weil formula. In this work, we \textrm{investigate} the physical
interpretation of these fluctuations in the bulk physics and pinpoint the
states giving rise to them. For this purpose, we \textrm{explore} an
ensemble of generalised Narain CFTs and \textrm{build} the AdS$_{3}$
gravitational dual in the Chern-Simons (CS) framework. We \textrm{study} the
associated charged BTZ black hole solution and \textrm{assess} its
stability. Using the Swampland weak gravity conjecture, we \textrm{show}
that the fluctuations of the ensemble average are below the BTZ threshold
and \textrm{correspond} to a sublattice of superextremal states emitted by
the black hole. We\textrm{\ exploit} the logarithmic density of states to
derive bounds on the charged vectors of the abelian CS symmetry and \textrm{%
introduce }a novel formulation of the density function to ensure consistency
with the sublattice WGC. We \textrm{establish }bounds that allows to\textrm{%
\ distinguish} heavy states contributing to the average from light states
generating fluctuations around it.
}


\begin{document}
\notoc
\maketitle
\flushbottom
\newpage
\tableofcontents
\section{Introduction}
\label{sec:intro}

The Swampland program \cite{SP1}-\cite{SP4} defines constraints on various
aspects of the effective field theories that can be completed into UV
quantum gravitational theories (Landscape models) such as the bound on the
rank of possible gauge groups \cite{Finit1}-\cite{Finit4}, the condition on
the mass of charged states \cite{WGC, Rajae3} and the structure of the
moduli space \cite{DC}. While the Swampland constraints are often backtested
in string theory which helps in motivating and formulating them, the
Swampland program is independent of a particular UV completion including
string theory.

Recent studies \cite{Land1}-\cite{Rajae2} have identified an intriguing
region within the Landscape, populated by 3D topological gravitational
theories, which has provided new insights into the Swampland program that
had not been apparent previously. This lead to establishing new connections
between the Swampland program and holography in quantum gravity research
based on quantum gravitational theories with exceptional boundaries given by
averaged ensembles \cite{AJ,MW}.

The usual AdS$_{3}$/CFT$_{2}$ correspondence associates a quantum gravity
theory in AdS$_{3}$ with a 2D CFT on its asymptote. However, recent
gravitational theories are shown to be dual not to a\ single boundary theory
but rather to an average over an ensemble of theories. This multi-theory
boundary was first observed and studied in JT gravity for AdS$_{2}$/CFT$_{1}$
\cite{JT}$,$ where the ensemble average is computed over a random matrix
theory \cite{JT1}. In 3D, the average is performed over an ensemble of
Narain CFTs; these models are well-known from toroidal compactifications,
there are theories of c free bosons with $U(1)_{L}^{c}\times U(1)_{R}^{c}$
current algebra where the positive integer $c$ defines the central charge ($%
c_{L}=c_{R}$). The average is generated by a bulk described by an exotic
theory of quantum gravity, where the sum over possible geometries is
captured by the abelian Chern-Simons (CS) theory $U(1)_{L}^{c}\times
U(1)_{R}^{c}$ \cite{AJ, MW}$.$ A generalization of Narain's CFTs can be
envisioned along the lines of \cite{M1}, where one can consider instead of
the even self dual lattices, general lattices with indefinite quadratic
forms. This defines an abelian CS theory with symmetry $U(1)_{L}^{p}\times
U(1)_{R}^{q}$ and $\Delta c=p-q$ central charge.

Recently, it has been demonstrated that these generalised Narain theories
exhibit global symmetries which is strictly prohibited by the Swampland
program. Specifically, the no global symmetries conjecture postulates that
all global symmetries of quantum gravitational theories must be either
broken or gauged \cite{NG}. Accordingly, to circumvent them, one must
consider fluctuations around the ensemble average, which correspond
mathematically to deviations in the Siegel-Weil formula \cite{SW1}-\cite{SW3}
or equivalently to the \textrm{Roeckle-Selberg} spectral decomposition
\textrm{\cite{RS1, RS2}}. However, it remains necessary to address how these
fluctuations might be interpreted in the bulk and how they arise from the
bulk to the boundary.

In this paper, we investigate the bulk equivalent of these fluctuations and
explore how additional Swampland criteria may impose constraints on the
averaged theory to ensure the consistency of the gravitational bulk. We
focus on the Weak Gravity Conjecture (WGC) \cite{wgc3, chark, MI} and
non-supersymmetric AdS conjecture \cite{Insta} to further probe the
relationship between ensemble average and the Swampland program. We consider
an ensemble of generalised Narain CFTs and build the corresponding AdS$_{3}$
gravitational dual in the Chern-Simons formulation by coupling the basic $%
SL\left( 2,\mathbb{R}\right) _{L}\times SL\left( 2,\mathbb{R}\right) _{R}$\
to abelian\textrm{\ }$U(1)^{p+q-2}$ gauge fields. We study the charged BTZ
black hole and examine its stability. To derive a constraint relation that
regulates the discharge of the black hole, we exploit the modular invariance
to compute the sublattice WGC. We show that light superextremal states below
the black hole threshold lead to fluctuations around the ensemble averaging.
As a consistency check, we employ the logarithmic density of states to
derive constraints on the charge vectors of the abelian CS symmetry. To
ensure physical consistency with the sublattice WGC, we propose an
alternative formulation of the density function by treating the bounds on
the charged states as permeable barriers. This provided a novel approach to
derive the WGC condition and allowed to categorize states into two \textrm{%
phases}: heavy (averaged states) and light (fluctuation states).

The organisation of this paper is as follows: In \autoref{sec2}, we begin by
describing the differences and similarities between the standard and
generalised Narain CFTs and establish the mathematical framework for
ensemble averaging. Then, we study the emergence of global symmetries and
address their removal by considering fluctuations around the ensemble
average. In \autoref{sec3}, we construct the AdS$_{3}$ gravitational bulk of the
ensemble average of generalised Narain CFTs. We consider Einstein AdS$_{3}$\
gravity in the Chern-Simons formulation coupled to U(1)$^{p+q-2}$\ gauge
fields. We show that under diagonal boundary conditions, the asymptotic
symmetries yield two copies of the asymptotic $U(1)_{L}^{p}\times
U(1)_{R}^{q}$\ affine algebra and the inaugural Brown-Henneaux boundary
conformal symmetry emerges as a composite invariance through a twisted
Sugawara construction. In \autoref{sec4}, we investigate the physical
interpretation of these fluctuations from the lens of the bulk physics and
identify the states giving rise to them. We start by imposing the modular
invariance of the partition function to determine the states populating the
sublattice verifying WGC. Then, we demonstrate that light superextremal
states generate the fluctuations using the torus density of states. \autoref{sec5} is reserved for final comments and conclusions.
\vspace{-0.5cm}
\section{Swampland program of averaged ensembles}
\label{sec2}
In this section, we revisit the 3D Landscape of consistent quantum
gravitational theories given by holographic models with averaged ensembles
as boundary dual. We study generalised Narain CFTs to formulate their
average over the generalised Narain moduli space. We examine the emerging
global symmetries G$_{global}$ and investigate the restoration of the
consistency of the gravitational theories in accordance with the Swampland
program forbidding\emph{\ }G$_{global}$ by considering fluctuations away
from the average of the partition function of the boundary defined by the
generalised CFTs.

\subsection{Averaged CFTs and the bulk dual}

\qquad With the aim of formulating the average of an ensemble of generalised
Narain's two dimensional conformal field theories ($\mathcal{GNCFT}$), we
begin by considering a set $\mathcal{CFT}_{\Pi _{p,p}}$ of Narain CFTs with
moduli space given by \cite{AJ, MW}:%
\begin{equation}
\mathcal{M}_{\Pi _{p,p}}=O\left( p,p,\mathbb{Z}\right) \backslash O\left(
p,p,\mathbb{R}\right) /\left[ O\left( p,\mathbb{R}\right) \times O\left( p,%
\mathbb{R}\right) \right]  \label{pi}
\end{equation}%
where $\Pi _{p,p}$ is \textrm{an} even self-dual lattice. Often, Narain
lattice CFTs are treated with regard to toroidal compactifications in string
theory where $\Pi _{p,p}$ is given by the lattice $\Gamma _{16+p,p}$
describing the compactification of 10D heterotic string on the p-dimensional
torus $\mathbb{T}^{p}$. The moduli space $\mathcal{M}_{\Gamma _{16+p,p}}$ is
well-defined and parameterised by the background metric $G$ and the two-form
field $B$. The Hilbert-Fock space $\boldsymbol{H}$ of the compactified
(10-p) supersymmetric heterotic string theory factorises in the light-cone
formalism as follows \cite{het0}:%
\begin{equation}
\boldsymbol{H}=\boldsymbol{H}_{B}^{8-p,8-p}\otimes \boldsymbol{H}%
_{F}^{0,8-p}\otimes \boldsymbol{H}_{\Pi _{p,p}}
\end{equation}%
The bosonic $\boldsymbol{H}_{B}^{8-p,8-p}$ and the fermionic subspaces built
out of the string oscillator modes $\alpha _{n}^{\mu },$ $\tilde{\alpha}%
_{n}^{\mu }$\ and $\tilde{\psi}_{n}^{\mu }$\ are restricted to the
transverse directions%
\begin{equation}
\begin{tabular}{|c|c|ccc|cc|ccc|}
\hline
{\small string\ modes} & \multicolumn{4}{|c|}{$\boldsymbol{H}_{B}^{8-p,8-p}$}
&  &  & \multicolumn{3}{|c|}{$\boldsymbol{H}_{F}^{0,8-p}$} \\ \hline
{\small left} &  & $\tilde{\alpha}_{n}^{i}$ & $,$ & $2\leq i\leq 9-p$ &  &
& $\tilde{\psi}_{r}^{i}$ &  & $2\leq i\leq 9$ \\ \hline
{\small right} &  & $\alpha _{n}^{i}$ & $,$ & $2\leq i\leq 9-p$ &  &  &  &
&  \\ \hline
\end{tabular}%
\end{equation}%
The vector space $\boldsymbol{H}_{\Gamma _{16+p,p}}$ labeled by even
self-dual lattice $\Gamma _{16+p,p}$ has the structure \cite{het}%
\begin{equation}
\boldsymbol{H}_{\Gamma _{16+p,p}}=\boldsymbol{H}_{B}^{16+p,0}\otimes \left(
\dbigoplus\limits_{\mathbf{\vec{p}}\in \Gamma _{16+p,p}}\mathbb{C}\left\vert
\mathbf{\vec{p}}\right\rangle \right)
\end{equation}%
where $\boldsymbol{H}_{B}^{16+p,0}$ is constructed by the oscillators $%
\alpha _{n}^{i}$ with $10-p\leq i\leq 26-p$; sometimes we refer to them like
$\beta _{n}^{\text{\textsc{a}}}$ with $1\leq $\textsc{a}$\leq 16+p.$

However, because we target a generalisation of these CFTs to study the
ensemble averages, we focus on a broader class of these theories without
being tied to specific lattice configurations and treat Narain models
independently of any string embedding for a simplified framework. As a
contribution, we will ensure to highlight the parallels, whenever feasible,
between the stringy Narain CFTs and the generalised CFTs described below.

A generalisation of the standard Narain CFTs can be envisioned in the lines
of \cite{M1}:%
\begin{equation}
\mathcal{M}_{Q^{{\small (p,q)}}}=\boldsymbol{O}_{\mathfrak{Q}^{{\small (p,q)}%
}}\left( \mathbb{Z}\right) \backslash \boldsymbol{O}\left( p,q,\mathbb{R}%
\right) /\left[ \boldsymbol{O}\left( p,\mathbb{R}\right) \times \boldsymbol{O%
}\left( q,\mathbb{R}\right) \right]  \label{ms}
\end{equation}%
where $\mathfrak{Q}^{{\small (p,q)}}$ is an indefinite quadratic form of
rank $p+q$ and signature $(p,q)$; it is characterised by a ${\small %
(p+q)\times (p+q)}$\textrm{\ }integer\textrm{\ }matrix $\mathrm{Q}_{\text{%
\textsc{ab}}}^{{\small (p,q)}}$ as given by the formal expansion
\begin{equation}
\mathfrak{Q}_{\boldsymbol{l,l}}^{{\small (p,q)}}=\sum_{\text{\textsc{a,b}}%
=1}^{p+q}l^{\text{\textsc{a}}}\mathrm{Q}_{\text{\textsc{ab}}}^{{\small (p,q)}%
}l^{\text{\textsc{b}}}\qquad ,\qquad \boldsymbol{l}%
=(l_{1},...,l_{p};l_{p+1},...,l_{p+q})  \label{form}
\end{equation}%
with integer vector $\boldsymbol{l}$ belonging to the lattice $\mathbb{Z}%
^{p+q}.$ The set $\boldsymbol{O}_{\mathfrak{Q}^{{\small (p,q)}}}\left(
\mathbb{Z}\right) $ is a subgroup of the\textrm{\ }orthogonal transformations%
\textrm{\ }$\mathcal{O}$\textrm{\ }in\textrm{\ }$\boldsymbol{O}\left( p,q,%
\mathbb{Z}\right) $ that preserves the quadratic form $\mathfrak{Q}_{%
\boldsymbol{l,l}}^{{\small (p,q)}}.$

\textbf{Bulk dual:} The bulk dual of the above generalised CFTs is given by
the three dimensional abelian $U(1)^{p+q}$ Chern-Simons theory. Its field
action is parameterised by the integral quadratic from $\mathrm{Q}^{{\small %
(p,q)}}$ as follows:%
\begin{equation}
S_{\text{\textsc{cs}}}\left[ A;\mathrm{Q}^{{\small (p,q)}}\right] =\frac{i}{%
8\pi }\int_{M_{3}}d^{3}x\sum_{\text{\textsc{ab}}=1}^{p+q}A^{\text{\textsc{a}}%
}\mathrm{Q}_{\text{\textsc{ab}}}^{{\small (p,q)}}dA^{\text{\textsc{b}}}
\label{CS1}
\end{equation}%
Because of the coupling matrix\textrm{\ }$\mathrm{Q}_{\text{\textsc{ab}}}^{%
{\small (p,q)}}$\textrm{,} the generalised Narain's moduli space $\mathcal{M}%
_{Q^{{\small (p,q)}}}$ includes two sectors given by\textrm{\ }\emph{even}
and \emph{odd} integral quadratic $\mathfrak{Q}_{\boldsymbol{l,l}}^{{\small %
(p,q)}}$ forms \cite{M1}. In this regard, recall that\textrm{\ }$\mathfrak{Q}%
_{\boldsymbol{l,l}}^{{\small (p,q)}}$\textrm{\ }is\textrm{\ }said to be even
(resp.\emph{\ odd}) if it is \emph{even} (resp.\emph{\ odd}) for any integer
vector $\boldsymbol{l}.$ In the interesting case\ of even $\mathfrak{Q}_{%
\boldsymbol{l,l}}^{{\small (p,q)}}$, the associated generalised partition
function $Z_{\mathfrak{Q}^{{\small (p,q)}}}\left( \boldsymbol{m},\tau ,%
\tilde{\tau}\right) $ for a point $\boldsymbol{m}$ of the moduli space $%
\mathcal{M}_{\mathfrak{Q}^{{\small (p,q)}}}$ is given by \cite{M3}:
\begin{equation}
Z_{Q}\left( \boldsymbol{m},\tau ,\tilde{\tau}\right) =\frac{\vartheta
_{Q}\left( \boldsymbol{m},\tau ,\tilde{\tau}\right) }{\eta \left( \tau
\right) ^{p}\tilde{\eta}\left( \tilde{\tau}\right) ^{q}}  \label{Z1}
\end{equation}%
where $\tau =\tau _{1}+i\tau _{1}$ is the modulus of the world sheet torus $%
\mathbb{T}^{2}$, $\eta \left( \tau \right) $ is the Dedekind eta function
and $\vartheta _{Q}\left( \boldsymbol{m},\tau ,\tilde{\tau}\right) $ is the
Siegel-Narain theta function.

The moduli space $\mathcal{M}_{\mathfrak{Q}}$ is locally homogeneous and
carries a Zamolodchikov metric \cite{Zchi}. For standard Narain CFTs with
stringy background, it can be formulated in terms of the metric $\mathbf{G}$
and two-form field $\mathbf{B}$\ as:%
\begin{equation}
ds^{2}=\mathbf{G}^{mk}\mathbf{G}^{nl}\left( d\mathbf{G}_{mn}d\mathbf{G}%
_{kl}+d\mathbf{B}_{mn}d\mathbf{B}_{kl}\right)
\end{equation}%
It provides a natural way to measure distances \textsc{d}$\left( \boldsymbol{%
m}_{1},\boldsymbol{m}_{2}\right) $ between different points $\boldsymbol{m}%
_{1}$ and $\boldsymbol{m}_{2}$ in the moduli space; it has a finite measure
for any $p,q>1$ \cite{MW}. This allows for the possibility to integrate over
the Narain moduli and define an averaged ensemble of CFTs as follows:%
\begin{equation}
\left\langle Z_{Q}\left( \boldsymbol{m},\tau ,\tilde{\tau}\right)
\right\rangle _{\mathcal{M}_{Q}}:=\frac{1}{vol\left( \mathcal{M}_{Q}\right) }%
\int_{\mathcal{M}_{Q}}\left[ d\boldsymbol{m}\right] Z_{Q}\left( \boldsymbol{m%
},\tau ,\tilde{\tau}\right)
\end{equation}%
where $\left\langle ...\right\rangle _{\mathcal{M}_{Q}}$ refers to averaging
over generalised Narain's moduli space $\mathcal{M}_{Q}$. And the
normalisation factor%
\begin{equation}
vol\left( \mathcal{M}_{Q}\right) :=\int_{\mathcal{M}_{Q}}\left[ d\boldsymbol{%
m}\right]
\end{equation}%
defines the volume of the moduli space. The $\left( \tau ,\bar{\tau}\right) $
denote the modular parameters of the genus-one Riemann surface $\Sigma
\simeq \mathbb{T}^{2}.$ Furthermore, the averaged Siegel-Narain theta
function $\Theta \left( \tau ,\tilde{\tau}\right) _{\mathcal{M}_{Q}}$ is
given by
\begin{equation}
\Theta \left( \tau ,\tilde{\tau}\right) _{\mathcal{M}_{Q}}:=\frac{1}{%
vol\left( \mathcal{M}_{Q}\right) }\int_{\mathcal{M}_{Q}}\left[ d\boldsymbol{m%
}\right] \vartheta _{Q}\left( \boldsymbol{m,}\tau ,\tilde{\tau}\right)
\end{equation}%
Thus, the averaged ensemble's partition function is:
\begin{equation}
\left\langle Z_{Q}\left( \boldsymbol{m,}\tau ,\tilde{\tau}\right)
\right\rangle _{\mathcal{M}_{Q}}=\frac{\left\langle \vartheta _{Q}\left(
\boldsymbol{m,}\tau ,\tilde{\tau}\right) \right\rangle _{\mathcal{M}_{Q}}}{%
\eta \left( \tau \right) ^{p}\tilde{\eta}\left( \tilde{\tau}\right) ^{q}}
\label{z1}
\end{equation}%
And following the Siegel-Weil theorem \cite{SW1,SW2,SW3}%
\begin{equation}
\left\langle \vartheta _{Q}\left( \boldsymbol{m,}\tau ,\tilde{\tau}\right)
\right\rangle _{\mathcal{M}_{Q}}=E_{Q}\left( \tau ,\tilde{\tau}\right)
\label{avev}
\end{equation}%
the partition function (\ref{z1}) can be put in the form
\begin{equation}
\left\langle Z_{Q}\left( \boldsymbol{m},\tau ,\tilde{\tau}\right)
\right\rangle _{\mathcal{M}_{Q}}=\frac{E_{Q}\left( \tau ,\tilde{\tau}\right)
}{\eta \left( \tau \right) ^{p}\tilde{\eta}\left( \tilde{\tau}\right) ^{q}}
\label{vz}
\end{equation}%
where $E_{Q}\left( \tau ,\tilde{\tau}\right) $\ is the Siegel-Eisenstein
series that can be interpreted as a sum over geometries in the gravitational
dual \cite{M2,PT}. This $E_{Q}\left( \tau ,\tilde{\tau}\right) $ is modular
invariant and can be presented like%
\begin{equation}
E_{Q}\left( \tau ,\tilde{\tau}\right) =\sum_{\gamma \in P\backslash SL\left(
2,\mathbb{Z}\right) }\func{Im}\left( \gamma \tau \right)
\end{equation}%
where $\gamma $ belongs to $P\backslash SL\left( 2,\mathbb{Z}\right) $ and
where%
\begin{equation}
P=\left\{ M=\left(
\begin{array}{cc}
1 & n \\
0 & 1%
\end{array}%
\right) ;\quad n\in \mathbb{Z}\right\} \qquad ;\qquad \tau ^{\prime }=\tau +n
\end{equation}%
It is a subgroup of $SL\left( 2,\mathbb{Z}\right) $ that leaves $\func{Im}%
\left( \tau \right) $ invariant.

Correspondingly, we associate the partition function of the averaged CFTs
ensemble $\left\langle Z_{Q}\left( \boldsymbol{m},\tau ,\tilde{\tau}\right)
\right\rangle _{\mathcal{M}_{Q}}$ to a sum over 3D\textrm{\ }geometries
\textsc{Y} in the bulk Chern-Simons theories $\sum_{\text{\textsc{Y}}}Z_{%
\text{\textsc{Y}}}^{U(1)^{p+q}}\left( \tau ,\tilde{\tau}\right) $ such that
\cite{PT}:%
\begin{equation}
\sum_{\text{handlebodies Y}}Z_{\text{\textsc{Y}}}^{U(1)^{p+q}}\left( \tau ,%
\tilde{\tau}\right) =\left\langle Z_{Q}\left( \boldsymbol{m},\tau ,\tilde{%
\tau}\right) \right\rangle _{\mathcal{M}_{Q}}  \label{ads/cft}
\end{equation}%
with $\tau $ denoting the modular parameters of the boundary surface $\Sigma
=\partial $\textsc{Y}. Topologically, the 3D geometries\textrm{\ \textsc{Y} }%
are solid torii, with 2-torus boundaries,\textrm{\ }known as genus- one
handlebodies. As detailed in \cite{MW}, the bulk CS theory (\ref{CS1}) does
not give any insights on how to choose \textsc{Y} with boundary $\partial $%
\textsc{Y} and the CS path integral should a priori hold for any geometry
\textsc{Y}. Nevertheless, in order to sum over \textsc{Y}, the Siegel-Weil
formula requires the boundary $\partial $\textsc{Y} to \emph{be connected};
and in this case, the simplest class of three dimensional topologies is
given by handlebodies.

Before concluding, let us discuss the \emph{odd} integral case of $\mathrm{Q}%
^{{\small (p,q)}}$. For this choice of quadratic forms, one can uncover a
relevant affiliation to the spin structure. The averaged partition function
is computed over fermionic CFTs where the boundary torus possesses four spin
structures stemming from the first Homology group as follows: $H^{1}\left(
\mathbb{T}^{2};\mathbb{Z}_{2}\right) =\mathbb{Z}_{2}\oplus \mathbb{Z}_{2}$.
We label the $\mathbb{Z}_{2}$\ signs by $\left( \epsilon _{1},\epsilon
_{2}\right) $. The Siegel-Weil formula is given by the Eisenstein series
with spin structure $\left( \epsilon _{1},\epsilon _{2}\right) $\ like \cite%
{M1}:%
\vspace{-0.1cm}
\begin{equation}
\left\langle \vartheta _{Q}^{\epsilon _{1},\epsilon _{2}}\left( \tau ,\tilde{%
\tau}\right) \right\rangle _{\mathcal{M}_{Q}}=E_{Q}^{\epsilon _{1},\epsilon
_{2}}\left( \tau ,\tilde{\tau}\right)
\end{equation}%
\vspace{-0.1cm}
The bulk dual in this instance is given by the spin Chern-Simons theory also
known as half integer level CS theories \cite{scs1}. The CS spin action is
usually defined by the extension on a bounding 4-manifold\textrm{\ }$M_{4}$
with boundary $M_{3}=\partial M_{4}$ as follows\textrm{:}%
\vspace{-0.1cm}
\begin{equation}
S_{CS}=\frac{i}{8\pi }\int_{M_{3}}A^{\text{\textsc{a}}}\mathrm{Q}_{\text{%
\textsc{ab}}}^{{\small (p,q)}}dA^{\text{\textsc{b}}}=2\pi i\int_{M_{4}}\frac{%
1}{8\pi ^{2}}\frac{dA^{\text{\textsc{a}}}\mathrm{Q}_{\text{\textsc{ab}}}^{%
{\small (p,q)}}dA^{\text{\textsc{b}}}}{2}
\end{equation}%
\vspace{-0.1cm}
To exhibit the dependence on the spin structure, let us consider the CS
functional $e^{S_{CS}}.$ It shows up in the path integral and must be
single-valued \cite{scs2}, meaning that $\int_{M_{4}}\frac{1}{8\pi ^{2}}%
\frac{dA^{\text{\textsc{a}}}\mathrm{Q}_{\text{\textsc{ab}}}^{{\small (p,q)}%
}dA^{\text{\textsc{b}}}}{2}$\ should be integral. For the case of even $%
\mathfrak{Q}^{{\small (p,q)}}$, this is guaranteed due to $\frac{1}{2}dA^{%
\text{\textsc{a}}}\mathrm{Q}_{\text{\textsc{ab}}}^{{\small (p,q)}}dA^{\text{%
\textsc{b}}}$\ being \emph{integral} and to $\int_{M_{4}}\frac{1}{8\pi ^{2}}%
\frac{dA^{\text{\textsc{a}}}\mathrm{Q}_{\text{\textsc{ab}}}^{{\small (p,q)}%
}dA^{\text{\textsc{b}}}}{2}$ being the first \textrm{P}ontryagin class which
is an integer element of the cohomology class. As for the case of odd $%
\mathfrak{Q}^{{\small (p,q)}},$ the term $\frac{1}{2}dA^{\text{\textsc{a}}}%
\mathrm{Q}_{\text{\textsc{ab}}}^{{\small (p,q)}}dA^{\text{\textsc{b}}}$ is
no longer integer. However, if we endow the manifold $M_{3}$ with a spin
structure, and by extension $M_{4}$ as well, the new functional is then
given by $e^{\frac{1}{2}S_{CS}}$. Therefore, the term $\int_{M_{4}}\frac{1}{%
8\pi ^{2}}\frac{dA^{\text{\textsc{a}}}\mathrm{Q}_{\text{\textsc{ab}}}^{%
{\small (p,q)}}dA^{\text{\textsc{b}}}}{2}$ becomes even integer since for
spin structured manifolds the first Pontryagin class is even and so the $e^{%
\frac{1}{2}S_{CS}}$ is single valued.\newline
In the subsequent sections, we disregard odd quadratic forms and focus the
analysis on the former even case.

In sum, we have established the difference between the standard Narain CFTs
and the generalised ones which can be recapitulated in the following table:%
\begin{equation}
\text{%
\begin{tabular}{|c||c|c|}
\hline
\textbf{Aspect} &
\begin{tabular}{c}
\textbf{Standard} \\
\textbf{Narain CFTs}%
\end{tabular}
&
\begin{tabular}{c}
\textbf{Generalized} \\
\textbf{Narain CFTs}%
\end{tabular}
\\ \hline\hline
\textbf{Lattice } &
\begin{tabular}{c}
{\small Unimodular even} \\
{\small self-dual lattice}%
\end{tabular}
&
\begin{tabular}{c}
{\small Indefinite} \\
{\small non-unimodular lattices}%
\end{tabular}
\\ \hline
\textbf{%
\begin{tabular}{c}
\textbf{Moduli} \\
\textbf{space}%
\end{tabular}%
} &
\begin{tabular}{c}
{\small Well-defined by the background} \\
{\small metric G and two-form B.}%
\end{tabular}%
{\small \ } & {\small Parameterised by }$\mathfrak{Q}^{{\small (p,q)}}$%
{\small \ } \\ \hline
\textbf{%
\begin{tabular}{c}
\textbf{Partition} \\
\textbf{function}%
\end{tabular}%
} & {\small Modular invariant} &
\begin{tabular}{c}
{\small Generally} \\
non modular invariant%
\end{tabular}
\\ \hline
\textbf{%
\begin{tabular}{c}
\textbf{String} \\
\textbf{theory}%
\end{tabular}%
} &
\begin{tabular}{c}
{\small Heterotic} \\
{\small string compactifications}%
\end{tabular}
&
\begin{tabular}{c}
{\small Not necessarily} \\
{\small tied to strings}%
\end{tabular}
\\ \hline
\begin{tabular}{c}
\textbf{Intended} \\
\textbf{applications}%
\end{tabular}%
\textbf{\ } &
\begin{tabular}{c}
{\small Target spaces }$\mathbb{T}^{p}$ \\
{\small with world sheet }$\mathbb{T}^{2}$%
\end{tabular}
&
\begin{tabular}{c}
{\small Ensemble of solid tori } \\
{\small geometries Y with }$\mathbb{T}^{2}${\small \ boundaries}%
\end{tabular}
\\ \hline
\end{tabular}%
}  \label{tab}
\end{equation}%
In the upcoming subsection, we study the consistency of bulk gravitational
theories with regard to the no-global symmetries of the Swampland program

\subsection{Swampland constraints on ensemble averaging}

\qquad From the previous subsection, one might ponder on the implications of
averaging over the generalised Narain moduli space on the bulk quantum
gravitational theories (\ref{ads/cft}). A major prospect was mentioned in
\cite{M1}, where averaging the ensemble of dual CFTs can serve as a path
integral formulation of the Euclidian bulk\ quantum gravity theory,
\begin{equation}
Z_{\text{\textsc{Y}}}\left[ \mathrm{\beta ,J}_{E}\right] =\dint\nolimits_{%
\text{\textsc{Y}}}\left[ Dg_{\mu \nu }\right] e^{-\frac{1}{\hbar }\mathcal{S}%
_{E}\left[ g_{\mu \nu },\mathrm{\beta ,J}_{E}\right] }
\end{equation}%
where $\mathrm{\beta }$ is the inverse of temperature ($\mathrm{\beta =1/T}$%
) and $\mathrm{J}_{E}$ the angular momentum. Averaging over the CFT moduli
space $\left\langle Z_{Q}\left( \boldsymbol{m},\tau ,\tilde{\tau}\right)
\right\rangle _{\mathcal{M}_{Q}}$ can be considered as an alternative to the
summing $\sum_{\text{\textsc{Y}}}Z_{\text{\textsc{Y}}}^{U(1)^{p+q}}$ over
all possible geometries to account for all possible configurations \textsc{Y}
of spacetime that could contribute to the gravitational
 dynamics.

This is a reflection of the duality associating the CFT moduli space $%
\mathcal{M}_{\mathfrak{Q}}$ with that of Riemann surfaces $\Sigma _{\text{%
\textsc{Y}}}=\partial $Y. It can be exhibited through the \emph{Howe duality}
of the \emph{reductive dual pairs} \cite{R1}-\cite{R3}. This later states
that a pair of groups $(G,H)$\ is a dual pair if they are both embedded in a
larger symplectic group $G^{\nparallel }$\ such that $G$\ is the centralizer
of $H$\ and vice versa. On the other hand, the Howe duality relates
irreducible representations of the two groups of the dual pair.
Particularly, the set of irreducible representation of the group $G$\ given
by $R\left( G\right) $\ are in a 1:1 correspondence with the set of
irreducible representation of the group $H$\ given by $R\left( H\right) .$\
For our setting, the dual pair is given by the CFT moduli space symmetry $%
G=O\left( p,q,\mathbb{R}\right) $\ and the mapping class group $H=Sp\left(
2g,\mathbb{R}\right) $\ of the 2D manifolds of genus $g$\ at the spacetime's
surface;\ they are both embedded inside the larger symplectic group $%
G^{\nparallel }=Sp\left( 2g\left( p+q\right) ,\mathbb{R}\right) $ \cite{M1}%
.\

In this subsection we will study other aspects of the relationship between
holography and quantum gravity with regard to the Swampland program \cite%
{SP1, SP2}, to shed more light on how can one contribute to the
understanding of the other. We will go over the analysis of \cite{M2},
studying the emergence of global symmetries \textrm{G}$_{{\small global}}$
and the intertwinement with the Swampland criteria, particularly the no
global symmetry \cite{NG} and the distance conjectures \cite{DC}. This study
can be viewed as a preparatory step for our investigation that will cover
additional conjectures, particularly the weak gravity \cite{WGC} and the
instability of non-supersymmetric AdS \cite{Insta}.

The Swampland program provides a set of consistency conditions that must be
verified in order for effective field theories coupled to gravity to be UV
completed. One of the key swampland criteria is the no- global symmetries
conjecture. It postulates that all global symmetries of quantum
gravitational theories must be either broken or gauged \cite{NG}. Most of
the arguments forbidding global symmetries stem from black hole physics. For
instance the No-hair theorem that states that a black hole (BH) can be
characterised by its mass M$_{BH}$, gauged charge\textrm{\ }Q$_{BH}$ and
angular momentum J$_{BH}$. Therefore, an additional global charge Q$%
_{global} $ would not figure in the black hole properties. And since the
horizon is blind to this global charge, $(\mathbf{i})$ the black hole
evaporation will not effect it and will end up as a remnant \cite{NH}.
Moreover, $(\mathbf{ii})$ we will not be able to determine this global
charge which gives in consequence an infinite uncertainty to an observer
outside the black hole corresponding to an infinite amount of entropy which
is of course a violation of the holographic principle expecting it to be
finite \cite{NG}.\newline
Now, are global symmetries present in the averaged theories?

In \cite{M2}, it has been demonstrated that an ensemble symmetry G
connecting different theories ---say two theories \QTR{cal}{T}$h$ and $%
\mathcal{T}h^{\prime }$ within a set of theories\textrm{\ }\{$\mathcal{T}%
h_{1},\mathcal{T}h_{2},...$\} related as $\mathcal{T}h^{\prime }=$ \textrm{g}%
$\mathcal{T}h\mathrm{g}^{-1}$--- can be turned into an emergent \emph{global}
symmetry \textrm{G}$_{{\small global}}$ of the averaged theory $\left\langle
\mathcal{T}h\right\rangle _{\mathcal{M}}$. To see this, let us consider the
ensemble average of an observable $\mathcal{O}\left( \boldsymbol{m},x\right)
$ over a moduli space $\mathcal{M}_{Q}$ as follows:%
\begin{equation}
\left\langle \mathcal{O}\left( x\right) \right\rangle _{\mathcal{M}_{Q}}:=%
\frac{1}{vol\left( \mathcal{M}_{Q}\right) }\int_{\mathcal{M}_{Q}}\left[ d%
\boldsymbol{m}\right] \mathcal{O}\left( \boldsymbol{m},x\right)
\end{equation}%
Assuming\ there exists a moduli space $\mathcal{M}_{Q}$ symmetry $G$ such
that we have\textrm{\ }$\left( \mathbf{i}\right) $ the invariance of the
measure under G namely $\left[ d\left( \boldsymbol{m}^{{\small (g)}}\right) %
\right] =\left[ d\boldsymbol{m}\right] $ with $\boldsymbol{m}^{{\small (g)}%
}=g.\boldsymbol{m},$ and $\left( \mathbf{ii}\right) $\ the symmetry\textrm{\
}$\mathcal{O}\left( \boldsymbol{m}^{{\small (g)}},x^{{\small (g)}}\right) =%
\mathcal{O}\left( \boldsymbol{m},x\right) ,$ then $G$ is an ensemble
symmetry that maps the theory ($\mathcal{T}h$) at the point $\boldsymbol{m}$
to another theory ($\mathcal{T}h^{\prime }$) at the transformed point $%
\boldsymbol{m}^{{\small (g)}}$; i.e:%
\begin{equation}
\mathcal{T}h\left[ \boldsymbol{m}^{{\small (g)}}\right] =\mathrm{g}\mathcal{T%
}h\left[ \boldsymbol{m}\right] \mathrm{g}^{-1}\qquad ,\qquad \mathrm{g}\in
\mathrm{G}
\end{equation}%
This ensemble symmetry implies the equality of the averaging of observables $%
\mathcal{O}\left( x\right) $ in $\mathcal{T}h\left[ \boldsymbol{m}\right] $
and their homologue $\mathcal{O}\left( g.x\right) $ in the theory $\mathcal{T%
}h\left[ \boldsymbol{m}^{{\small (g)}}\right] ,$%
\begin{equation}
\left\langle \mathcal{O}\left( g.x\right) \right\rangle _{\mathcal{M}%
_{Q}}=\left\langle \mathcal{O}\left( x\right) \right\rangle _{\mathcal{M}%
_{Q}}
\end{equation}%
therefore making $G$ a global symmetry of a single theory $\mathcal{T}h_{0}$
since the dependence on the moduli space variables $\boldsymbol{m}$ is
removed. By\textrm{\ }using variation language, this global feature can be
formally stated as follows
\begin{equation}
\frac{\delta }{\delta \boldsymbol{m}}\mathcal{T}h_{0}\sim 0\qquad
\Rightarrow \qquad G=\mathrm{G}_{{\small global}}
\end{equation}

This is not the only way the global symmetries can emerge in effective field
theories; they can also follow from taking the infinite distance limit in
the moduli space [$\lim $\textsc{d}$_{\mathcal{M}}\left( \boldsymbol{m}_{1},%
\boldsymbol{m}_{2}\right) \rightarrow \infty $]. In fact the Swampland
distance conjecture stipulates that at the asymptote of the moduli space $%
\mathcal{M}_{{\small asymptote}}$, where we often expect gauge couplings
going to zero [$g_{YM}^{{\small asymptote}}\rightarrow 0$], an infinite
tower of states becomes exponentially light \cite{DC}:%
\begin{equation}
\text{\textsc{m}}\left( \mathbf{p}\right) \sim \text{\textsc{m}}_{0}e^{-%
\mathrm{\alpha }\text{\textsc{d}}_{\mathcal{M}}\left( \mathbf{p}\right)
}\qquad ,\qquad \boldsymbol{m}:=\mathbf{p}
\end{equation}%
with $\mathrm{\alpha }$ \textrm{some} positive constant. In this relation,
the \textsc{m}$\left( \mathbf{p}\right) $ is the mass at a point $\mathbf{p}$
in the moduli space $\mathcal{M}_{Q}$ and \textsc{m}$_{0}$\ is a reference
mass scale. The \textsc{d}$_{\mathcal{M}}\left( \mathbf{p}\right) $ is the
geodesic distance between the reference point $\mathbf{p}_{0}$ and a generic
point $\mathbf{p}$. For \textsc{d}$_{\mathcal{M}}\left( \mathbf{p}\right)
\rightarrow \infty ,$ we have \textsc{m}$\left( \mathbf{p}\right)
\rightarrow 0$. In this asymptotic limit, the gauge symmetry\textrm{\ G}$%
_{gauge}$\textrm{\ }gets replaced by a global symmetry\textrm{\ }$\mathrm{G}%
_{{\small global}}$\textrm{\ }for which\textrm{\ }$g_{YM}^{{\small asymptote}%
}\rightarrow 0$\textrm{.}

Therefore, global symmetries \textrm{G}$_{{\small global}}$ can emerge $%
\left( \mathbf{i}\right) $ either by considering the average of the CFTs
ensemble; or $\left( \mathbf{ii}\right) $ by taking the infinite distance
limit of the generalised Narain's moduli space. Accordingly, in order for
the gravitational bulk theories dual to the ensemble averaging to be
consistent, the boundary global symmetries \textrm{G}$_{{\small global}}$
must be broken down\textrm{:}
\begin{equation}
\mathrm{G}_{{\small global}}\qquad \rightarrow \qquad \{I_{id}\}
\end{equation}%
\textrm{\ }This can be done through two equivalent ways as described in \cite%
{M2}. Either one must settle the effective theory at \emph{finite} distances
in the moduli space [\textsc{d}$\left( \mathbf{p}\right) <\infty $] for
which $g_{YM}\neq \left( g_{YM}^{{\small asymptote}}\rightarrow 0\right) $.
Or consider fluctuations away from the average (\ref{avev}) through
deviations $f\left( \tau ,\tilde{\tau}\right) $ measured via the \textrm{%
Roeckle-Selberg} spectral decomposition \cite{RS1, RS2} as follows:%
\begin{equation}
f\left( \tau ,\tilde{\tau}\right) =\tau _{2}^{(p+q)/4}\delta \vartheta _{Q}
\label{Fluc}
\end{equation}%
where:%
\begin{equation}
\delta \vartheta _{Q}=\left\langle \vartheta _{Q}\left( \tau ,\tilde{\tau}%
\right) \right\rangle _{\mathcal{M}_{Q}}-E_{Q}\left( \tau ,\tilde{\tau}%
\right)
\end{equation}%
If we take into consideration the relation linking the imaginary component
of the torus $\tau _{2}$\ with the inverse of temperature $\tau _{2}=1/T,$\
the fluctuations become temperature dependent. In fact large fluctuations $%
f\left( \tau ,\tilde{\tau}\right) >>0$ correspond to $\tau _{2}\rightarrow
\infty ,$\ i.e low temperatures $T\rightarrow 0.$\ Inversely, at high
temperatures $T\rightarrow \infty $, we get soft fluctuations $f\left( \tau ,%
\tilde{\tau}\right) \rightarrow 0$. Therefore, the fluctuations average out
at high temperatures where $\Theta \left( \tau ,\tilde{\tau}\right) _{%
\mathcal{M}_{Q}}$ is superimposed with the quantity $E_{Q}\left( \tau ,%
\tilde{\tau}\right) $. We will discuss this in more details in terms of the
density of states in \textbf{subsection 4.2}.

This intrinsic tie linking ensemble averages with the Swampland program
might be the key to enhance our holographic understanding of the Swampland
conjectures. The criteria on the other hand may help us to put constraints
on the averaged ensemble and its gravitational dual. Yet, it also raises
doubts on whether this link is confined to specific Swampland constraints
like the no global symmetries conjecture and should not be generalised for
all the Swampland program criteria. Before delving into this point, we shall
first build a gravitational dual given by Einstein gravity coupled to
Abelian CS gauge fields in this next section to introduce cosmological
solutions such as BTZ black holes.

\section{Gravitational dual of averaged $\mathcal{GNCFT}$'s}
\label{sec3}
\qquad The 3D bulk abelian $U(1)^{p+q}$ Chern-simons theory of the averaged
CFTs is only an approximation that had been successful in reproducing many
results of the yet to be established 3D gravitational theory \textrm{\cite%
{MW}, \cite{Ofer}-\cite{G2}}. An in-depth description of the nature of the
bulk falls outside the scope of this paper; instead we will :

\begin{description}
\item[$\left( i\right) $] think about the values of this 3D abelian $%
U(1)^{p+q}$ gauge CS theory in terms of an \emph{asymptotic} effective
diagonal description of a larger non abelian CS theory with potential\ $%
A_{\mu }\left( \mathbf{x}\right) =\sum_{\text{\textsc{b}}}A_{\mu }^{\text{%
\textsc{b}}}\left( \mathbf{x}\right) \boldsymbol{t}_{\text{\textsc{b}}}$ and
bulk gauge symmetry $G_{gauge}=G_{L}\times G_{R}$ containing $U(1)^{p+q}$ as
a subsymmetry; and

\item[$\left( ii\right) $] look for diagonal boundary solutions $A_{\mu }^{%
{\small (bnd)}}\left( \mathbf{\sigma }\right) =\sum_{\text{\textsc{b}}%
}\delta _{\mu }^{\alpha }\boldsymbol{a}_{\alpha }^{\text{\textsc{b}}}\left(
\mathbf{\sigma }\right) \boldsymbol{\hat{Q}}_{\text{\textsc{b}}}$ solving
AdS/CFT conditions that can be roughly imagined in terms of the asymptotic
limit of the bulk potential as follows\textrm{\ }%
\begin{equation}
\lim_{\rho \rightarrow \infty }A_{\mu }\left( \mathbf{x}\right) =A_{\mu }^{%
{\small (bnd)}}\left( \mathbf{\sigma }\right) \qquad ,\qquad \mathbf{x}%
=(t;\rho ,\varphi )\qquad ,\qquad \mathbf{\sigma }=(t;\varphi )  \label{lim}
\end{equation}%
with $\mathbf{\sigma }$ parameterising the boundary surface (here a
cylinder) and $A_{\mu }^{{\small (bnd)}}=\delta _{\mu }^{\alpha }\boldsymbol{%
a}_{\alpha }\left( \mathbf{\sigma }\right) +b^{-1}\partial _{\mu }b$ as well
as $A_{\rho }^{{\small (bnd)}}=b^{-1}\partial b/\partial \rho $ with gauge
transformation $b$. Schematically, we have%
\begin{equation}
\begin{tabular}{l|ll}
fields$\diagdown $ spacetime & \quad bulk $(t;\rho ,\varphi )$ & \quad
boundary $(t;\varphi )$ \\ \hline
gauge symmetry & \quad $G_{L}\times G_{R}$\quad & \quad $U(1)_{L}^{p}\times
U(1)_{R}^{q}$\quad \\
potentials & \quad $A_{\mu L}\oplus A_{\mu R}$\quad & \quad $\boldsymbol{a}%
_{\alpha L}\oplus \boldsymbol{a}_{\alpha R}$ \\ \hline
\end{tabular}%
\end{equation}
\end{description}

In this regard, we notice that it has been argued in \cite{PT} that an AdS$%
_{3}\ $gravitational dual of the averaged CFT can be found provided that the
boundary potential living on\textrm{\ }$\partial $AdS$_{3}$ sits in the
diagonal subgroup $U(1)^{p+q}$ of the asymptotic gauge symmetry for the case
where $p=q$. Our proposal goes in this direction; and consequently it
constitutes for us a basic algorithm in the construction of the
gravitational dual of averaged $\mathcal{GNCFT}$s with moduli space as in eq(%
\ref{ms}) with $p\neq q$.

\subsection{Realising the asymptotic limit (\protect\ref{lim})}

To deal with the asymptotic limit (\ref{lim}), we think about it in terms of
two boundary conditions $\left( \mathbf{BC1}\right) $ and $\left( \mathbf{BC2%
}\right) $.

First, the boundary potential $A_{\mu }^{{\small (bnd)}}$ is related to bulk
field $A_{\mu }=\sum_{\text{\textsc{b}}}A_{\mu }^{\text{\textsc{b}}}%
\boldsymbol{t}_{\text{\textsc{b}}}$ by a gauge transformation with group
elements $b=b\left( \mathbf{x}\right) $ as
\begin{equation}
\left( \mathbf{BC1}\right) :\qquad
\begin{tabular}{lll}
$A_{\mu }^{{\small (bnd)}}$ & $=$ & $b^{-1}A_{\mu }b+b^{-1}\partial _{\mu }b$
\\
$A_{\rho }^{{\small (bnd)}}$ & $=$ & $b^{-1}\frac{\partial }{\partial \rho }%
b $%
\end{tabular}
\label{bnd1}
\end{equation}%
where the radial component $A_{\rho }^{{\small (bnd)}}$ is a pure gauge
transformation; this implies that the 3 components of $A_{\mu }^{{\small %
(bnd)}}$ can be split as $(A_{\rho }^{{\small (bnd)}},\boldsymbol{a}_{\alpha
})$ with $\boldsymbol{a}_{\alpha }=(\boldsymbol{a}_{t},\boldsymbol{a}%
_{\varphi })$ and radial component $\boldsymbol{a}_{\rho }=0$.

Second, the boundary field $A_{\mu }^{{\small (bnd)}}$ is required to be a
diagonal potential; that is commuting gauge fields $[A_{\mu }^{{\small (bnd)}%
},A_{\nu }^{{\small (bnd)}}]=0.$ As such it must be valued in the Cartan
subalgebra of the gauge symmetry group G$_{gauge}.$
\begin{equation}
\left( \mathbf{BC2}\right) :\qquad
\begin{tabular}{lll}
$A_{\mu }^{{\small (bnd)}}$ & $=$ & $\dsum\limits_{\text{\textsc{b}}}%
\boldsymbol{a}_{\mu }^{\text{\textsc{b}}}\left( \sigma \right) \boldsymbol{%
\hat{Q}}_{\text{\textsc{b}}}$ \\
$A_{\rho }^{{\small (bnd)}}$ & $=$ & $b^{-1}\frac{\partial }{\partial \rho }%
b $%
\end{tabular}
\label{bnd2}
\end{equation}%
with \textsc{Q}$_{\text{\textsc{b}}}$'s obeying%
\begin{equation}
\left[ \boldsymbol{\hat{Q}}_{\text{\textsc{b}}},\boldsymbol{\hat{Q}}_{\text{%
\textsc{c}}}\right] =0
\end{equation}%
they generate the asymptotic boundary gauge symmetry $U(1)^{p+q}$. Notice
also the following features:

\begin{description}
\item[A)] the boundary conditions $\left( \mathbf{BC1}\right) $ and $\left(
\mathbf{BC2}\right) $ put a strong constraint on the allowed values of the
group elements $b(\mathbf{x}).$ As example, we give this transformation for
the bulk gauge sub-symmetry $SL\left( 2,\mathbb{R}\right) _{L}\times
SL\left( 2,\mathbb{R}\right) _{R}$ contained in $G_{gauge};$ it has the form%
\textrm{\ }\cite{Rajae2}%
\begin{equation}
b(\mathbf{x})=e^{-\frac{1}{\zeta }L_{+}}e^{-\frac{1}{2}\rho L_{-}}\qquad
,\qquad \tilde{b}(\mathbf{x})=e^{+\frac{1}{\zeta }L_{+}}e^{+\frac{1}{2}\rho
L_{-}}  \label{b}
\end{equation}%
with $L_{0,\pm }$ being the generators of $SL\left( 2,\mathbb{R}\right) $
satisfying $[L_{n},L_{m}]=\left( n-m\right) L_{n+m}.$

\item[B)] Because of boundary conformal invariance, the two components of
boundary potentials $\boldsymbol{a}_{\alpha }=\sum \boldsymbol{a}_{\alpha }^{%
\text{\textsc{b}}}\boldsymbol{\hat{Q}}_{\text{\textsc{b}}}$ with label $%
\alpha =t,\sigma $ must be formulated in terms of conserved affine\ chiral
and anti-chiral abelian currents $\boldsymbol{j}_{z}$ and $\boldsymbol{\bar{j%
}}_{\bar{z}}.$ This means that the natural frame to deal with the boundary
potential is given by $\sigma ^{\pm }=\varphi \pm t.$ As such, the local
dependence of $\boldsymbol{a}_{\alpha }$ decomposes like%
\begin{equation}
\boldsymbol{a}_{\alpha }=\boldsymbol{a}_{+}\oplus \boldsymbol{a}_{-}
\end{equation}%
with%
\begin{equation}
\begin{tabular}{lll}
$\frac{\partial }{\partial \sigma ^{-}}\boldsymbol{a}_{+}$ & $=$ & $0$ \\
$\frac{\partial }{\partial \sigma ^{+}}\boldsymbol{a}_{-}$ & $=$ & $0$%
\end{tabular}%
\qquad \Rightarrow \qquad
\begin{tabular}{lll}
$\boldsymbol{a}_{+}$ & $=$ & $\boldsymbol{a}_{+}\left( \sigma ^{+}\right) $
\\
$\boldsymbol{a}_{-}$ & $=$ & $\boldsymbol{a}_{-}\left( \sigma ^{-}\right) $%
\end{tabular}%
\end{equation}%
These features imply that $\boldsymbol{\hat{Q}}$ decomposes in turn into
left and right components like $\left( \boldsymbol{Q}_{L},\boldsymbol{Q}%
_{R}\right) $ with%
\begin{equation}
\boldsymbol{Q}_{L}=\left( \boldsymbol{Q}_{i}\right) _{1\leq i\leq p}\qquad
,\qquad \boldsymbol{Q}_{R}=(\boldsymbol{\tilde{Q}}_{\bar{\imath}})_{1\leq
\bar{\imath}\leq q}
\end{equation}%
where the $\boldsymbol{Q}_{L}$ generates the asymptotic $U(1)_{L}^{p}$ and $%
\boldsymbol{Q}_{R}$ concerns the boundary $U(1)_{R}^{q}$.
\end{description}

In this regard, it is important to note that when considering this diagonal
abelian boundary symmetry as the product of two distinct left and right
group sectors, we must distinguish between two arising situations:

\begin{description}
\item[$\left( \mathbf{i}\right) $] theories with\textrm{\ }$p=q=D$\textrm{\ }%
having an exact left/right symmetry $\boldsymbol{Q}_{L}\leftrightarrow
\boldsymbol{Q}_{R}$. This case can be associated with Narain CFTs whose
moduli space\textrm{\ }is given by (\ref{pi}). It permits to split (\ref%
{bnd2})\ as follows%
\begin{equation}
\begin{tabular}{lllll}
$\boldsymbol{a}_{+}$ & $=$ & $\dsum\limits_{i=1}^{D}\boldsymbol{a}_{+}^{i}%
\boldsymbol{Q}_{L_{i}}$ & $\equiv $ & $\dsum\limits_{i=1}^{D}\boldsymbol{a}%
_{+}^{i}\boldsymbol{Q}_{i}$ \\
$\boldsymbol{\tilde{a}}_{-}$ & $=$ & $\dsum\limits_{i=1}^{D}\boldsymbol{%
\tilde{a}}_{-}^{\bar{\imath}}\boldsymbol{Q}_{R\bar{\imath}}$ & $\equiv $ & $%
\dsum\limits_{i=1}^{D}\boldsymbol{\tilde{a}}_{-}^{\bar{\imath}}\boldsymbol{%
\tilde{Q}}_{\bar{\imath}}$%
\end{tabular}%
\end{equation}%
By performing the Wick rotation $t=it_{E},$ the $\sigma ^{+}$ is mapped to
the complex $z=\varphi +it_{E}$ and $\boldsymbol{a}_{+}$ gets replaced by $%
\boldsymbol{a}_{z}$. Similarly, the $\sigma ^{-}$\ gets mapped into its
complex conjugate $\bar{z}$ and $\boldsymbol{\tilde{a}}_{-}$ will be
replaced by $\boldsymbol{\bar{a}}_{\bar{z}}.$ Moreover, denoting by $c_{L}$
and $c_{R}$ the conformal anomalies in the left and right sectors, the
difference
\begin{equation}
\Delta c=\left\vert c_{L}-c_{R}\right\vert \qquad ,\qquad \Delta c=0
\label{dc}
\end{equation}%
vanishes as the theories possess equal right and left degrees of freedom.
Below, we think about this vanishing value as a consistency condition.

\item[$\left( \mathbf{ii}\right) $] Conformal models with $p\neq q$ and $%
D=(p+q)/2$ where left and right sectors are dissymmetric having non
vanishing $\Delta c=\left\vert c_{L}-c_{R}\right\vert $; thus violating (\ref%
{dc}). This is an anomalous defect which has to be addressed. Notice that in
this situation, we either have\textrm{\ }$p<q$ or $p>q$ for which $\Delta
=(p-q)/2$ is non vanishing. To overcome this difficulty, one may either
cancel the anomalies using the inflow mechanism \cite{Finit2, Finit4} or
consider adding fermions to reinstate $\Delta c_{tot}=0$. Following the
later, we ensure the validity of (\ref{dc}) by adding (left or right)
fermions to restore balance between $c_{L}$\ and $c_{R}$. \textrm{For }$%
D=(p+q)/2$\textrm{\ to be an integer rank,} $p$\ and $q$\ must have \emph{%
the same parity}; that is both even or both odd. These are the theories
associated with the generalised Narain CFTs\ with generalised moduli space
as in (\ref{ms}). Here, the diagonal constraint ($\mathbf{BC2}$) allows the
boundary potential (\ref{bnd2}) to expand as follows%
\begin{equation}
\begin{tabular}{lll}
$\boldsymbol{a}_{z}$ & $=$ & $\dsum\limits_{i=1}^{p}\boldsymbol{Q}%
_{i}a_{z}^{i}\left( z\right) $ \\
$\boldsymbol{a}_{\bar{z}}$ & $=$ & $\dsum\limits_{\bar{\imath}=1}^{q}%
\boldsymbol{\bar{Q}}_{\bar{\imath}}\bar{a}_{\bar{z}}^{\bar{\imath}}\left(
\bar{z}\right) $%
\end{tabular}
\label{zpz}
\end{equation}%
with
\begin{equation}
\frac{\partial }{\partial \bar{z}}a_{z}^{i}=0\qquad ,\qquad \frac{\partial }{%
\partial z}\bar{a}_{\bar{z}}^{\bar{\imath}}=0
\end{equation}%
We solve these boundary conditions in terms of compactified 2D conformal
with scalar and chiral fermi fields meaning that $a_{z}^{i}$\ and $\bar{a}_{%
\bar{z}}^{\bar{\imath}}$\ have realisations in terms of scalars $\left(
X^{i},\bar{Y}^{\bar{\imath}}\right) $\ and fermions\textrm{\ }($\psi _{L},%
\bar{\chi}_{R}$). As illustration, we consider the case $p<q$ with $%
D=(p+q)/2 $ integer and set $q=p+2n$; we have the splitting
\begin{equation}
a_{z}=\left( a_{z}\right) _{scalar}+\left( a_{z}\right) _{fermi}\qquad
,\qquad \bar{a}_{\bar{z}}=\left( \bar{a}_{\bar{z}}\right) _{scalar}
\end{equation}%
with%
\begin{equation}
\begin{tabular}{lllll}
$\left( a_{z}^{i}\right) _{scalar}$ & $=$ & $\frac{\partial }{\partial z}%
X^{i}$ & $\qquad ,\qquad $ & $1\leq i\leq p$ \\
$\left( a_{z}^{\theta }\right) _{fermi}$ & $=$ & $\bar{\psi}_{z/2}^{\theta
}\psi _{z/2}^{\theta }$ & $\qquad ,\qquad $ & $1\leq \theta \leq 2n$ \\
$\left( \bar{a}_{\bar{z}}^{\bar{\imath}}\right) _{scalar}$ & $=$ & $\frac{%
\partial }{\partial \bar{z}}\bar{Y}^{\bar{\imath}}$ & $\qquad ,\qquad $ & $%
1\leq \bar{\imath}\leq q+2n$%
\end{tabular}
\label{th}
\end{equation}%
and
\begin{equation}
\frac{\partial }{\partial \bar{z}}\frac{\partial }{\partial z}X^{i}=0\qquad
,\qquad \frac{\partial }{\partial z}\psi _{z/2}^{\theta }=0\qquad ,\qquad
\frac{\partial }{\partial z}\frac{\partial }{\partial \bar{z}}\bar{Y}^{\bar{%
\imath}}=0
\end{equation}%
giving the free field equations of $X^{i},$ $\psi _{z/2}^{\theta },\bar{\psi}%
_{z/2}^{\theta }$\ and $\bar{Y}^{\bar{\imath}}.$ Subsequently, we conceal
the fermionic contribution and focus only on the scalar field contributions.
\end{description}

\subsection{3D gravity coupled to $p+q-2$ abelian potentials}

\textrm{\ }An interesting field representation of the boundary constraint (%
\ref{bnd1}-\ref{bnd2}) is given by the spin s=2 Einstein gravity on AdS$_{3}$
coupled to the 3d gauge fields $A_{L}\left( \mathbf{x}\right) \oplus
A_{R}\left( \mathbf{x}\right) $ valued in the Lie algebra of the abelian
gauge symmetry%
\begin{equation}
G_{abelian}=U(1)_{L}^{p-1}\times U(1)_{R}^{q-1}
\end{equation}%
In this subsection, we will go over this new set up, derive the resulting
asymptotic algebra $\mathcal{G}_{asymp}$ and study the solutions to the
equations of motion of AdS$_{3}$\ gravity coupled to the\textrm{\ }$%
G_{abelian}$ \textrm{CS}\ gauge radiations.

We begin by considering the ordinary spin $s=2$ Einstein gravity on \textsc{%
Y=}AdS$_{3}$ coupled to $U(1)^{p-1}\times U(1)^{q-1}$ gauge fields with 3D
field action given by%
\begin{equation}
\begin{tabular}{lll}
$\mathcal{S}[g_{\mu \nu },A_{\mu },\tilde{A}_{\mu }]$ & $=$ & $\frac{1}{%
16\pi G_{N}}\dint\nolimits_{\text{\textsc{Y}}}\sqrt{-g}\left( R-2\Lambda
\right) $ \\
&  & $+\dsum\limits_{i=1}^{p-1}\frac{1}{4\pi }\dint\nolimits_{\text{\textsc{Y%
}}}A^{i}\mathrm{Q}_{ij}dA^{j}$ \\
&  & $-\dsum\limits_{a=1}^{q-1}\frac{1}{4\pi }\dint\nolimits_{\text{\textsc{Y%
}}}\tilde{A}^{a}\mathrm{\tilde{Q}}_{ab}d\tilde{A}^{b}$%
\end{tabular}
\label{ECS}
\end{equation}%
where%
\begin{equation}
\mathrm{Q}_{ij}=\kappa _{i}\delta _{ij}\qquad ,\qquad \mathrm{\tilde{Q}}%
_{ab}=\tilde{\kappa}_{a}\delta _{ab}
\end{equation}%
with $\kappa _{i}$ and $\tilde{\kappa}_{a}$\ CS levels. We intend to develop
the investigation of (\ref{ECS}) in the Chern-Simons framework \cite{AT, W}
where spin s=2 gravity is described by two sl$(2,\mathbb{R})$ gauge fields
denoted by the 1-form potentials $A_{{\small sl}_{{\small 2}}}$ and $\tilde{A%
}_{{\small sl}_{{\small 2}}}$. In this description, the two sl$(2,\mathbb{R}%
) $ potentials are related to the so(1,2) spin connection form $\mathbf{%
\omega }$\ and the Dreibein $\mathbf{e}$ as follows:%
\begin{equation}
A_{{\small sl}_{{\small 2}}}=\mathbf{\omega }+\frac{1}{l_{{\small AdS}}}%
\mathbf{e}\qquad ,\qquad \tilde{A}_{{\small sl}_{{\small 2}}}=\mathbf{\omega
}-\frac{1}{l_{{\small AdS}}}\mathbf{e}
\end{equation}%
In terms of these 1-form left and right Chern-Simons potentials $A_{{\small %
sl}_{{\small 2}}}$ and $\tilde{A}_{{\small sl}_{{\small 2}}}$, the pure
gravity field action $\mathcal{S}_{{\small grav}}$ takes the form
\begin{equation}
\begin{tabular}{lll}
$\frac{1}{16\pi G_{N}}\dint\nolimits_{Y}\sqrt{-g}\left( R-2\Lambda \right) $
& $=$ & $+\frac{\kappa _{0}}{4\pi }\dint\nolimits_{Y}tr\left( A_{{\small sl}%
_{{\small 2}}}dA_{{\small sl}_{{\small 2}}}+\frac{2}{3}A_{{\small sl}_{%
{\small 2}}}^{3}\right) $ \\
&  & $-\frac{\kappa _{0}}{4\pi }\dint\nolimits_{Y}tr\left( \tilde{A}_{%
{\small sl}_{{\small 2}}}d\tilde{A}+\frac{2}{3}\tilde{A}_{{\small sl}_{%
{\small 2}}}^{3}\right) $%
\end{tabular}
\label{rt}
\end{equation}%
with 3D Newton constant $G_{N}$ and Chern Simons level $\kappa _{0}$ linked
to one another and to $l_{{\small AdS}}$ as follows
\begin{equation}
\kappa _{0}=\frac{l_{{\small AdS}_{{\small 3}}}}{4G_{N}}\qquad ,\qquad
\Lambda =-\frac{1}{l_{{\small AdS}_{{\small 3}}}^{2}}
\end{equation}%
The resulting gauge theory with total action $CS$ given by eqs(\ref{ECS},\ref%
{rt}) can be therefore expressed as the difference of two CS theories $%
CS_{L}-CS_{R}$ described by distinct left and right potentials $\boldsymbol{A%
}_{L}$, $\boldsymbol{A}_{R}$ splitting as
\begin{equation}
\boldsymbol{A}_{L}=A_{{\small sl}_{{\small 2}}}\oplus \sum_{i=1}^{p-1}A^{i}%
\boldsymbol{Q}_{i}\qquad ,\qquad \boldsymbol{A}_{R}=\tilde{A}_{{\small sl}_{%
{\small 2}}}\oplus \sum_{a=1}^{q-1}\tilde{A}^{a}\boldsymbol{\tilde{Q}}_{a}
\label{ALR}
\end{equation}%
The left-handed $CS_{L}$ has a non abelian gauge symmetry $G_{L}=SL(2,%
\mathbb{R})_{L}\times U(1)^{p-1}$ while the right-handed $CS_{R}$ has a
different non abelian gauge symmetry $G_{R}=SL(2,\mathbb{R})_{R}\times
U(1)^{q-1}$. To deal with the boundary constraint relations (\ref{bnd1}-\ref%
{bnd2}), we use as bulk coordinates the spherical $x^{\mu }=(t,\rho ,\varphi
)$ and as boundary surface ones, corresponding to the asymptotic limit $\rho
\rightarrow \rho _{\infty },$ the variables $\sigma ^{\alpha
}=(t_{E},\varphi )$. Here t$_{E}$ is the euclidian time component and $%
\varphi $ an angular variable taking values in $\left[ 0,2\pi \right[ .$

Asymptotically ($\rho \rightarrow \rho _{\infty }$), the bulk CS gauge field
$A_{\mu }=\left( A_{\rho },A_{\alpha }\right) $ with $A_{\alpha }=\left(
A_{t};A_{\varphi }\right) $ and its fellow $\tilde{A}_{\mu }=(\tilde{A}%
_{\rho },\tilde{A}_{\alpha })$ with $\tilde{A}_{\alpha }=(\tilde{A}_{t};%
\tilde{A}_{\varphi })$ take the boundary values%
\begin{equation}
\begin{tabular}{lllll}
$A_{\alpha }\left( t,\rho ,\varphi \right) $ & $\qquad \rightarrow \qquad $
& $A_{\alpha }^{{\small (bnd)}}$ & $=$ & $a_{\alpha }\left( \sigma
^{+},\sigma ^{-}\right) $ \\
$A_{\rho }\left( t,\rho ,\varphi \right) $ & $\qquad \rightarrow \qquad $ & $%
A_{\rho }^{{\small (bnd)}}$ & $=$ & $a_{\rho }=cte$%
\end{tabular}
\label{BC}
\end{equation}%
and%
\begin{equation}
\begin{tabular}{lllll}
$\tilde{A}_{\alpha }\left( t,\rho ,\varphi \right) $ & $\qquad \rightarrow
\qquad $ & $\tilde{A}_{\alpha }^{{\small (bnd)}}$ & $=$ & $\tilde{a}_{\alpha
}\left( \sigma ^{+},\sigma ^{-}\right) $ \\
$\tilde{A}_{\rho }\left( t,\rho ,\varphi \right) $ & $\qquad \rightarrow
\qquad $ & $\tilde{A}_{\rho }^{{\small (bnd)}}$ & $=$ & $\tilde{a}_{\rho
}=cte$%
\end{tabular}
\label{CB}
\end{equation}%
\begin{equation*}
\end{equation*}%
with boundary potentials $a_{\alpha }$ and $\tilde{a}_{\alpha }$ function
only of the chiral and antichiral variables $\sigma ^{\pm }=\varphi \pm t/l_{%
{\small AdS}}$ taken as in (\ref{zpz}). By using the Wick rotation $%
t=it_{E}, $ the variable $\sigma ^{+}$ gets mapped to the holomorphic $%
z=\varphi +it_{E}/l_{{\small AdS}}$ and the $\sigma ^{-}$ to its complex
conjugate $\bar{z}=\varphi -it_{E}/l_{{\small AdS}}.$

Applying the gauge transformation (\ref{bnd1}-\ref{bnd2}) to the $SL(2,%
\mathbb{R})^{2}\times U(1)^{p+q}$ Chern-Simons gauge theory $CS_{L}-CS_{R}$,
we have the asymptotic left and right boundary potentials%
\begin{equation}
A^{{\small (bnd)}}=b^{-1}\left( d+a\right) b\qquad ,\qquad \tilde{A}^{%
{\small (bnd)}}=\tilde{b}^{-1}\left( d+\tilde{a}\right) \tilde{b}
\end{equation}%
with $b\left( \mathbf{x}\right) $ and $\tilde{b}\left( \mathbf{x}\right) $
some gauge transformations of $sl(2,\mathbb{R})\times \widetilde{sl}(2,%
\mathbb{R});$ given by eq(\ref{b})$.$ Notice that as far as the boundary
conditions of the CS potentials are concerned, one has several possible
choices; $\left( \mathbf{i}\right) $ the Grumiller-Riegler gauge where $%
b=e^{L_{-}}e^{\rho L_{0}}$ with L$_{0,\pm }$ generating $sl(2,\mathbb{R})$
satisfying the commutation relation $\left[ L_{n},L_{m}\right] =\left(
n-m\right) L_{n+m}.$ $\left( \mathbf{ii}\right) $ the Ba\~{n}ados choice $%
b=e^{\rho L_{0}}$ or $\left( \mathbf{iii}\right) $ \textrm{t}he non
spherical gauge\textrm{\ }$b=$ $e^{-L_{+}/\zeta _{-}}e^{-\rho L_{-}/2}$ with
constant $\zeta _{-}$; for more details see \cite{Rajae2,Y}. The choices $%
\left( \mathbf{i}\right) $ and $\left( \mathbf{ii}\right) $ lead to
non-diagonal boundary gauge potentials; they are not considered here.

For the bulk gauge fields (\ref{ALR}),\textrm{\ }the boundary potentials $A^{%
{\small (bnd)}}$ and $\tilde{A}^{{\small (bnd)}}$\ are chosen to obey the
diagonal boundary conditions with 1-form potentials $a=a_{z}dz$ and $\tilde{a%
}=\tilde{a}_{\bar{z}}d\bar{z}$ valued in the Cartan subalgebra of the gauge
symmetry:%
\begin{equation}
a\left( z\right) \quad {\Large \in }\quad u(1)_{L}\oplus u(1)_{L}^{\oplus
(p-1)}\qquad ,\qquad \tilde{a}\left( \bar{z}\right) \quad {\Large \in }\quad
u(1)_{R}\oplus u(1)_{R}^{\oplus (q-1)}
\end{equation}%
Explicitly, they are expressed as \cite{Soft1, Soft12, Soft2}:
\begin{equation}
a_{z}=\mathcal{J}_{z}^{0}L_{0}+\sum_{i=1}^{p-1}\mathcal{J}_{z}^{i}%
\boldsymbol{Q}\text{$_{i}$}\qquad ,\qquad \tilde{a}_{\bar{z}}=\mathcal{%
\tilde{J}}_{\bar{z}}^{0}L_{0}+\sum_{i=1}^{q-1}\mathcal{\tilde{J}}_{\bar{z}%
}^{i}\boldsymbol{\tilde{Q}}\text{$_{i}$}  \label{bc}
\end{equation}%
where \{$\boldsymbol{Q}_{i}$\} and \{$\boldsymbol{\tilde{Q}}_{i}$\} generate
$u(1)^{\oplus (q-1)}$ and $u(1)^{\oplus (p-1)}$ respectively. In these
relations, the functions $\left( \mathcal{J}_{z}^{0},\mathcal{J}%
_{z}^{i}\right) $ and $(\mathcal{\tilde{J}}_{\bar{z}}^{0},\mathcal{\tilde{J}}%
_{\bar{z}}^{i})$ are conserved (anti) chiral currents expanding in Laurent
modes like%
\begin{equation}
\begin{tabular}{lll}
$\mathcal{J}_{z}^{0}$ & $=$ & $\dsum\limits_{m\in \mathbb{Z}}z^{-m-1}J_{m}$
\\
$\mathcal{J}_{z}^{i}$ & $=$ & $\dsum\limits_{m\in \mathbb{Z}%
}^{m+1}z^{-m-1}J_{m}^{i}$%
\end{tabular}%
\qquad ,\qquad
\begin{tabular}{lll}
$\mathcal{\bar{J}}_{\bar{z}}^{0}$ & $=$ & $\dsum\limits_{m\in \mathbb{Z}}%
\bar{z}^{-m-1}\bar{J}_{m}$ \\
$\mathcal{\bar{J}}_{\bar{z}}^{i}$ & $=$ & $\dsum\limits_{m\in \mathbb{Z}%
}^{m+1}\bar{z}^{-m-1}\bar{J}_{m}^{i}$%
\end{tabular}%
\end{equation}%
Following the derivation of the asymptotic algebra algorithm, this yields
two copies of the asymptotic $U(1)_{L}^{p}\times U(1)_{R}^{q}$ affine
algebra extending the U$(1)^{2D}$ of \cite{PT}. It is generated by the
conserved charges ($J_{m}$, $\bar{J}_{m}$) and given as follows:%
\begin{equation}
i\left\{ J_{n}^{i},J_{m}^{j}\right\} =\frac{\kappa _{i}}{2}n\delta
^{ij}\delta _{m+n,0}\qquad ,\qquad i\left\{ \bar{J}_{n}^{a},\bar{J}%
_{m}^{b}\right\} =\frac{\tilde{\kappa}_{a}}{2}n\delta ^{ab}\delta _{m+n,0}
\label{asy}
\end{equation}%
where $i,j=0,1...p-1$ and $a,b=0,1...q-1$. And similarly to standard Narain
CFTs, the currents are formed from compacts bosons as $J^{i}\left( z\right)
:=\partial X^{i}$\ and $\tilde{J}^{\bar{\imath}}\left( \bar{z}\right) :=\bar{%
\partial}\bar{Y}^{\bar{\imath}}$ as well as\textrm{\ }$J^{a}\left( z\right) =%
\bar{\psi}_{z/2}^{\theta }\psi _{z/2}^{\theta }$\textrm{\ for} $p<q$; see
also eq(\ref{th}). Also, it must be noted that the inaugural Brown-Henneaux
boundary conformal symmetry emerges as a composite invariance through a
twisted Sugawara construction \cite{Soft1}-\cite{BH}\textrm{.}

Therefore, the boundary conditions (\ref{bc}) with symmetries $\boldsymbol{A}%
_{{\small bnd}}^{\prime }=h_{{\small bnd}}^{-1}\left( d+\boldsymbol{A}_{%
{\small bnd}}\right) h_{{\small bnd}}$\textrm{\ }are implied by the
gravitational bulk after performing gauge transformations valued in $%
G_{L}\times G_{R}$ as shown below,
\begin{equation}
\begin{tabular}{ccc}
bulk &  & boundary \\ \hline\hline
&  &  \\
$\boldsymbol{A}\left( \mathbf{x}\right) $ & $\underrightarrow{\text{ \ \ \ \
}b\left( \mathbf{x}\right) \text{\ \ \ \ \ }}$ & $\boldsymbol{A}_{{\small bnd%
}}\left( \mathbf{\sigma }\right) =b^{-1}\left( d+\boldsymbol{a}\right) b$ \\
&  &  \\
$\downarrow h\left( \mathbf{x}\right) $ &  & $\downarrow h_{{\small bnd}%
}\left( \mathbf{\sigma }\right) $ \\
&  &  \\
$\boldsymbol{A}^{\prime }=h^{-1}\left( d+\boldsymbol{A}\right) h$ & $\quad
\underrightarrow{\text{ \ \ \ \ }b^{\prime }\left( \mathbf{x}\right) \text{\
\ \ \ \ }}\quad $ & $\boldsymbol{A}_{{\small bnd}}^{\prime }=h_{{\small bnd}%
}^{-1}\left( d+\boldsymbol{A}_{{\small bnd}}\right) h_{{\small bnd}}$ \\
&  &  \\ \hline
\end{tabular}%
\end{equation}%
Using the boundary condition $\boldsymbol{A}_{bnd}=b^{-1}\left( d+%
\boldsymbol{a}\right) b$ and its image $\boldsymbol{A}_{bnd}^{\prime
}=b^{\prime -1}\left( d+\boldsymbol{a}^{\prime }\right) b^{\prime }$ under
the bulk group element $h\left( \mathbf{x}\right) $, we learn from this
above diagram the relationship $h_{bnd}=b^{-1}hb^{\prime }.$

Moreover using (\ref{ECS},\ref{rt}), the bulk action reads as follows%
\begin{equation}
\begin{tabular}{lll}
$\mathcal{S}[g_{\mu \nu },A_{\mu },\tilde{A}_{\mu }]$ & $=$ & $\frac{\kappa
}{4\pi }\dint\nolimits_{\text{\textsc{Y}}}tr\left( A_{{\small sl}_{{\small 2}%
}}dA_{{\small sl}_{{\small 2}}}+\frac{2}{3}A_{{\small sl}_{{\small 2}%
}}^{3}\right) $ \\
&  & $+\dsum\limits_{i=1}^{p-1}\frac{1}{4\pi }\dint\nolimits_{\text{\textsc{Y%
}}}A^{i}\mathrm{Q}_{ij}dA^{j}$ \\
&  & $+\frac{\kappa }{4\pi }\dint\nolimits_{\text{\textsc{Y}}}tr\left( -%
\tilde{A}_{{\small sl}_{{\small 2}}}d\tilde{A}-\frac{2}{3}\tilde{A}_{{\small %
sl}_{{\small 2}}}^{3}\right) $ \\
&  & $+\dsum\limits_{a=1}^{q-1}\frac{1}{4\pi }\dint\nolimits_{\text{\textsc{Y%
}}}\tilde{A}^{a}\mathrm{\tilde{Q}}_{ab}d\tilde{A}^{b}$%
\end{tabular}
\label{GG}
\end{equation}%
This 3D field action can be presented in a condensed form by exploiting the
properties of the gauge symmetry
\begin{equation*}
\left[ sl(2,\mathbb{R})_{L}\times U(1)_{L}^{p-1}\right] \times \left[ sl(2,%
\mathbb{R})_{R}\times U(1)_{R}^{q-1}\right]
\end{equation*}%
In fact, using the 1-form potentials $\boldsymbol{A}=A_{{\small sl}_{{\small %
2}}}\oplus A_{{\small u(1)}^{p-1}}$ and $\boldsymbol{\tilde{A}}=\tilde{A}_{%
{\small sl}_{{\small 2}}}\oplus \tilde{A}_{{\small u(1)}^{q-1}},$ the field
action becomes:
\begin{equation}
\begin{tabular}{lll}
$\mathcal{S}[g_{\mu \nu },\boldsymbol{A}_{\mu },\boldsymbol{\tilde{A}}_{\mu
}]$ & $=$ & $\frac{1}{4\pi }\dint\nolimits_{\text{\textsc{Y}}%
}\dsum\limits_{i,j=0}^{p-1}\mathrm{Q}_{ij}tr\left( \boldsymbol{A}^{i}d%
\boldsymbol{A}^{j}+\frac{2}{3}\boldsymbol{A}^{i}\wedge \boldsymbol{A}%
^{j}\wedge \boldsymbol{A}\right) $ \\
&  & $+\frac{1}{4\pi }\dint\nolimits_{\text{\textsc{Y}}}\dsum%
\limits_{a,b=0}^{q-1}\mathrm{\tilde{Q}}_{ab}tr\left( \boldsymbol{\tilde{A}}%
^{a}d\boldsymbol{\tilde{A}}^{b}+\frac{2}{3}\boldsymbol{\tilde{A}}^{a}\wedge
\boldsymbol{\tilde{A}}^{b}\wedge \boldsymbol{A}\right) $%
\end{tabular}
\label{Sfin}
\end{equation}%
where
\begin{equation}
\mathrm{Q}_{ij}=\kappa _{i}\delta _{ij}\qquad ,\qquad \mathrm{\tilde{Q}}%
_{ab}=-\tilde{\kappa}_{a}\delta _{ab}
\end{equation}%
with $\tilde{\kappa}_{0}=\kappa _{0}.$

\subsection{Partition function and cosmological solution}

Based on the previous analysis we draw a significant conclusion, the
asymptotic limit of the Einstein-Chern-Simons theory in the Chern-Simons
formulation (\ref{Sfin}) with diagonal boundary conditions is given by an
averaged free 2D CFT over generalised Narain lattice. This can be exhibited
through an AdS/CFT correspondence-like relation where $A_{sl_{2}}^{bnd}$ is
expressed in the diagonal basis as follows \cite{PT}:%
\begin{eqnarray}
\sum_{\text{\textsc{Y}}\in \mathbb{J}}Z_{\text{\textsc{Y}}}^{SL(2,\mathbb{R}%
)^{2}\times U(1)^{p+q-2}}\left( \tau ,\tilde{\tau}\right) &=&\sum_{\text{%
\textsc{Y}}\in \mathbb{J}}\dint\nolimits_{\text{\textsc{Y}}}\left[ Dg_{\mu
\nu }\right] \prod_{\text{\textsc{l}}=1}^{p+q}\left[ DA_{\mu \text{\textsc{l}%
}}\right] e^{-\mathcal{S}_{E}[g_{\mu \nu },A_{\mu \text{\textsc{l}}}]} \\
&=&\sum_{\text{\textsc{Y}}\in \mathbb{J}}Z_{\text{\textsc{Y}}%
}^{U(1)^{p+q}}\left( \tau ,\tilde{\tau}\right)  \label{Z2} \\
&=&\left\langle Z_{\Sigma =\partial \text{\textsc{Y}}}\left( m,\tau ,\tilde{%
\tau}\right) \right\rangle _{\mathcal{M}_{Q}}
\end{eqnarray}%
with $\mathbb{J}$\ defining the set of different topologies of thermal
\textsc{Y} invariant under $SL\left( 2,\mathbb{Z}\right) /P$ such that $P$
is the subgroup generated by the translation $\tau \rightarrow \tau +1.$

Furthermore, regarding the solutions to the equations of motion, the CS
terms in the Einstein-Chern-Simons gravity theory do not couple to the
metric and hence the Einstein equations with a negative cosmological
constant remain as in the pure $SL(2,\mathbb{R})\times SL(2,\mathbb{R})$
gravity case. The most general solution is given by \cite{PT}:
\begin{equation}
ds^{2}=dr^{2}+\frac{l^{2}}{4}\left( \mathcal{J}^{2}dx^{2}+\mathcal{\tilde{J}}%
^{2}d\widetilde{x}^{2}-2\cosh \left( \frac{2r}{l}\right) \mathcal{J\tilde{J}}%
dxd\tilde{x}\right)  \label{bh}
\end{equation}%
and therefore still contains the generic BTZ black hole metric \cite{BTZ} as
a part of the spectrum. However, for our setting, the black hole is now
charged under $U(1)^{p-1}\times U(1)^{q-1}.$\newline
In the upcoming section, we will discuss the\textrm{\ }thermodynamical
peculiarities of AdS$_{D}$ black holes with a focus on the BTZ black holes
of AdS$_{3}$ using the dual ensemble averaging and Swampland principles.

\section{BTZ threshold and WGC on fluctuating ensemble average}
\label{sec4}
\qquad So far, we have established two main results: $\left( \mathbf{i}%
\right) $ in order to break the emerging global symmetries G$_{{\small global%
}}$ of the averaged ensemble, one must consider fluctuations $f$ away from
the ensemble\textrm{\ }average with deviations given by (\ref{Fluc}). And
that $\left( \mathbf{ii}\right) $ there exists a gravitational bulk dual
provided by Einstein-Chern-Simons theory (\ref{ECS}) with charged BTZ black
hole solution under the $U(1)^{p-1}\times U(1)^{q-1}$ symmetry.

Our aim for this section is two folded: First, we study the decay of the BTZ
black hole and its discharge in accordance with the instability of non
supersymmetric AdS and the weak gravity conjectures using modular invariance
considered in\textrm{\ \cite{MI}.} Then, in light of the decaying black hole
solution, we investigate the bulk physics giving rise to the fluctuations
around the ensemble average and the significance of the deviations (\ref%
{Fluc}). It turns out that they are induced by the coupling of topological
gravity to abelian Chern-Simons potentials with boundary conditions sourced
by chemical potentials. As a consistency check; we end the section by
analysing the torus density of states with regard to the averaged ensemble
to identify the fluctuations states labeled by the charges of $U(1)^{p+q-2}$.

\subsection{Constraints from modular invariance}

According to the non supersymmetric AdS conjecture, any non supersymmetric
AdS space---as well as locally lookalike AdS geometries--- must exhibit some
sort of instability. Given the isometry linking the BTZ black hole to the AdS%
$_{3}$ geometry, it is intriguing to see if the BTZ black hole is truly
unstable. In \cite{wgc3}, one distinguishes two types of black holes in AdS
spaces; large AdS black holes that are stable as they are in equilibrium
with their thermal bath, and unstable small black holes with mass $M_{BH}$
in need of radiating some of their charge $\boldsymbol{Q}_{BH}\in
U(1)^{p-1}\times U(1)^{q-1}$. The Swampland weak gravity conjecture regulate
their discharge by requiring the emission of super-extremal particles state $%
|$\textsc{m}$,\boldsymbol{q}>$ with a mass \textsc{m} to charge $\boldsymbol{%
q}$ ratio as:%
\begin{equation}
\left\Vert \boldsymbol{q}\right\Vert \quad \geq \quad \frac{\text{\textsc{m}}%
}{M_{BH}}\left\Vert \boldsymbol{Q}_{BH}\right\Vert
\end{equation}

Assuming that our BTZ black hole carrying a $U(1)^{p-1}\times U(1)^{q-1}$
charge has a small horizon in comparison with the AdS$_{3}$ radius ($%
r_{BH}<<l_{AdS_{3}}$), it is therefore unstable and in need of discharging.
The weak gravity constraint governing its decay was computed in \cite{wgc3,
MI} and we will briefly review its computation subsequently.

The derivation is based on requiring the dual CFT to be modular invariant
which is usual for CFTs dual to gravitational theories on threefolds Y.
Actually, the AdS/CFT prescription instructs us to sum over every 3D
geometry Y with the same asymptotic behavior which correspond to considering
a boundary with a torus topology. Generally, in AdS/CFT, the partition
function $Z_{CFT}\left( \tau ,\tilde{\tau};\mathbf{\mu },\mathbf{\tilde{\mu}}%
\right) $ includes black hole states in a SL(2,$\mathbb{Z}$) multiplet and
it is modular invariant; here $\mathbf{\mu }=\left( \mu _{1},...,\mu
_{p-1}\right) $ and $\mathbf{\tilde{\mu}}=\left( \tilde{\mu}_{1},...,\tilde{%
\mu}_{q-1}\right) $ are fugacities associated with extra $%
U(1)_{L}^{p-1}\times U(1)_{R}^{q-1}$ charges. The additional U(1)$^{p+q-2}$
sector is also modular invariant which makes the charge lattice spanned by
the charge vectors $\boldsymbol{q}$ even and \emph{self-dual} \cite{pol1,
pol2}.

For our setting, as we have already established, see table (\ref{tab}), the
genus-one partition function associated with generalised Narain CFTs are
generally non-modular invariant. This is due to the random choice of the
quadratic form $\mathfrak{Q}^{p,q}$\ corresponding to non-self dual
lattices. To restore the modular invariance, we can either $(\mathbf{i})$\
impose that the even integer lattices have to be self-dual like Niemeier
lattices for instance \cite{M1}. Or $(\mathbf{ii})$\ think about the
generalised CFT as a building block of a more encompassing CFT with a
modular invariant partition function. For instance, one can combine a
generalised Narain CFT with quadratic form $\mathfrak{Q}_{\boldsymbol{l,l}}^{%
{\small (p,q)}}=\sum_{\text{\textsc{a,b}}=1}^{p+q}l^{\text{\textsc{a}}}%
\mathrm{Q}_{\text{\textsc{ab}}}^{{\small (p,q)}}l^{\text{\textsc{b}}}$\ with
a "conjugate" one having\textrm{\ }$\mathfrak{\bar{Q}}_{\boldsymbol{l,l}}^{%
{\small (p,q)}}=-\mathfrak{Q}_{\boldsymbol{l,l}}^{{\small (p,q)}}$\textrm{\ }%
form \cite{M3}. In fact, this last proposal is more natural to us as it is
automatically fulfilled by virtue of building the gravitational AdS$_{%
\mathrm{3}}$\ dual with the two left and right sectors based respectively on
$\mathrm{Q}$ and $\mathrm{\tilde{Q}}$\textrm{, }see (\ref{Sfin}) with $%
\mathrm{Q}_{\text{\textsc{ab}}}^{{\small (p,q)}}$ and $\mathrm{\tilde{Q}}_{%
\text{\textsc{ab}}}^{{\small (p,q)}}$, up to $O(p,q)$\ transformations.%
\textrm{\ }It can be thought of as
\begin{equation}
\mathrm{Q}_{\text{\textsc{ab}}}^{{\small (p,q)}}\sim \kappa _{\text{\textsc{a%
}}}\eta _{\text{\textsc{ab}}}\qquad ,\qquad \mathrm{\tilde{Q}}_{\text{%
\textsc{ab}}}^{{\small (p,q)}}\sim -\tilde{\kappa}_{\text{\textsc{a}}}\eta _{%
\text{\textsc{ab}}}
\end{equation}%
where here $\eta _{\text{\textsc{ab}}}=diag(+p,-q)$ obeys $O^{T}\eta O=\eta $
with $O$ being an orthogonal transformation; see\textrm{\ \cite{class} }for
equivalent realisations.

The associated genus-one partition function $Z_{Y}^{SL(2,\mathbb{R}%
)^{2}\times U(1)^{p+q-2}}$ of the generalised CFT (\ref{Z2}) with symmetry $%
SL(2,\mathbb{R})^{2}\times U(1)^{p+q-2}$, at the tree level is given by:

\begin{equation}
Z_{Y}\left( \tau ,\tilde{\tau};\mathbf{\mu },\mathbf{\ \tilde{\mu}}\right)
=Tr\left[ q^{h-\frac{c}{24}}y^{\boldsymbol{Q}}q^{-\overline{h}-\frac{c}{24}}%
\tilde{y}^{\boldsymbol{\tilde{Q}}}\right]  \label{ZB}
\end{equation}%
where the $\boldsymbol{Q}\in U(1)^{p-1},$ $\boldsymbol{\tilde{Q}}\in $ $%
U(1)^{q-1}$ are the charges and $y^{\boldsymbol{Q}}=exp(2\pi i\mathbf{\mu .}%
\boldsymbol{Q})$ and $\tilde{y}^{\boldsymbol{\tilde{Q}}}=exp(-2\pi i\mathbf{%
\tilde{\mu}.}\boldsymbol{\tilde{Q}})$ are the associated fugacities with
chemical potential vectors $\mathbf{\mu },$\textbf{\ }$\mathbf{\tilde{\mu}}.$%
\ A priori, as discussed in length in \cite{MI}, it is not obvious how this
flavored partition function transforms under modular transformations.
However, it has been proved that it obeys the following universal rules:%
\begin{eqnarray}
Z_{Y}\left( \tau ,\tilde{\tau};\mathbf{\mu },\mathbf{\ \tilde{\mu}}\right)
&=&Z_{Y}\left( \tau +1,\tilde{\tau}+1;\mathbf{\mu },\mathbf{\ \tilde{\mu}}%
\right)  \label{ZY2} \\
Z_{Y}\left( \tau ,\tilde{\tau};\mathbf{\mu },\mathbf{\ \tilde{\mu}}\right)
&=&e^{-\frac{i\pi }{\tau }\mathbf{\mu }^{2}+\frac{i\pi }{\tilde{\tau}}%
\mathbf{\tilde{\mu}}^{2}}Z_{Y}\left( -\frac{1}{\tau },-\frac{1}{\tilde{\tau}}%
;\frac{\mathbf{\mu }}{\tau },\mathbf{\ }\frac{\mathbf{\tilde{\mu}}}{\tilde{%
\tau}}\right) \\
Z_{Y}\left( \tau ,\tilde{\tau};\mathbf{\mu },\mathbf{\ \tilde{\mu}}\right)
&=&Z_{Y}\left( \tau ,\tilde{\tau};\mathbf{\mu }+\rho ,\mathbf{\ \tilde{\mu}}+%
\tilde{\rho}\right) \qquad \forall \left( \rho ,\tilde{\rho}\right) \in
\Gamma  \label{ZY4}
\end{eqnarray}%
where $\Gamma \cong \mathbb{Z}^{p+q-2}$ is the period lattice. Combining the
above formulas, we get:%
\begin{equation}
Z_{Y}\left( \tau ,\tilde{\tau};\mathbf{\mu }+\tau \mathbf{\rho },\mathbf{\
\tilde{\mu}}+\tilde{\tau}\tilde{\rho}\right) =e^{-2i\pi \mathbf{\mu }\rho
-i\pi \rho ^{2}\tau +2i\pi \mathbf{\tilde{\mu}}\tilde{\rho}-i\pi \tilde{\rho}%
^{2}\tilde{\tau}}Z_{Y}\left( \tau ,\tilde{\tau};\mathbf{\mu },\mathbf{\
\tilde{\mu}}\right)
\end{equation}%
To see this, we start from (\ref{ZB}), and compute\textrm{\ }$Z_{Y}\left(
\tau ,\tilde{\tau};\mathbf{\mu }+\tau \mathbf{\rho },\mathbf{\ \tilde{\mu}}+%
\tilde{\tau}\tilde{\rho}\right) $ while using (\ref{ZY2}-\ref{ZY4}), we get%
\begin{eqnarray}
&&e^{-\left( \frac{\pi i}{\tau }\mu ^{2}+i\pi \tau \rho ^{2}+2i\pi \mu \rho
\right) +\left( \frac{\pi i}{\tau }\tilde{\mu}^{2}+i\pi \tilde{\tau}\tilde{%
\rho}^{2}+2i\pi \tilde{\mu}\tilde{\rho}\right) }\times Z_{Y}\left( -\frac{1}{%
\tau },-\frac{1}{\tilde{\tau}};\frac{\mathbf{\mu }+\tau \mathbf{\rho }}{\tau
},\mathbf{\ }\frac{\mathbf{\tilde{\mu}}+\tilde{\tau}\tilde{\rho}}{\tilde{\tau%
}}\right)  \notag \\
&=&e^{-2\pi i\mu \rho -\pi i\rho ^{2}\tau +2\pi i\tilde{\mu}\tilde{\rho}+\pi
i\tilde{\rho}^{2}\tilde{\tau}}\times Z_{Y}\left( \tau ,\tilde{\tau};\mathbf{%
\mu },\mathbf{\ \tilde{\mu}}\right)
\end{eqnarray}%
By putting\textrm{\ }$\hat{h}$ $\equiv h-\frac{1}{2}\boldsymbol{Q}^{2}$ and $%
\widehat{\tilde{h}}\equiv \tilde{h}-\frac{1}{2}\boldsymbol{\tilde{Q}}^{2}$,
we express (\ref{ZB}) like%
\begin{equation}
Z_{Y}\left( \tau ,\tilde{\tau};\mathbf{\mu },\mathbf{\ \tilde{\mu}}\right)
=Tr\left[ \left( q^{\hat{h}-\frac{c}{24}}q^{\frac{1}{2}\boldsymbol{Q}^{2}}y^{%
\boldsymbol{Q}}\right) \left( \tilde{q}^{-\widehat{\tilde{h}}-\frac{c}{24}}%
\tilde{q}^{\frac{1}{2}\boldsymbol{\tilde{Q}}^{2}}\tilde{y}^{\boldsymbol{%
\tilde{Q}}}\right) \right]
\end{equation}%
then using (\ref{ZY2}-\ref{ZY4}), we end up with
\begin{eqnarray}
&&e^{+2\pi i\mu \rho +\pi i\rho ^{2}\tau -2\pi i\tilde{\mu}\tilde{\rho}-\pi i%
\tilde{\rho}^{2}\tilde{\tau}}\times Z_{Y}\left( \tau ,\tilde{\tau};\mathbf{%
\mu }+\tau \mathbf{\rho },\mathbf{\ \tilde{\mu}}+\tilde{\tau}\tilde{\rho}%
\right)  \notag \\
&=&e^{+2\pi i\mu \rho +\pi i\rho ^{2}\tau -2\pi i\tilde{\mu}\tilde{\rho}-\pi
i\tilde{\rho}^{2}\tilde{\tau}}Tr\left[ \left( q^{\hat{h}-\frac{c}{24}}q^{%
\frac{1}{2}\boldsymbol{Q}^{2}}q^{\boldsymbol{Q.}\rho }y^{\mathbf{Q}}\right)
\left( \tilde{q}^{-\widehat{\tilde{h}}-\frac{c}{24}}\tilde{q}^{\frac{1}{2}%
\boldsymbol{\tilde{Q}}^{2}}\tilde{y}^{\boldsymbol{\tilde{Q}}}\tilde{q}^{%
\boldsymbol{\tilde{Q}}.\tilde{\rho}}\right) \right]
\end{eqnarray}%
and then%
\begin{eqnarray}
Z_{Y}\left( \tau ,\tilde{\tau};\mathbf{\mu },\mathbf{\ \tilde{\mu}}\right)
&=&Tr\left[ \left( q^{\hat{h}-\frac{c}{24}}q^{\frac{1}{2}\boldsymbol{Q}%
^{2}}y^{\boldsymbol{Q}}\right) \left( \tilde{q}^{-\widehat{\tilde{h}}-\frac{c%
}{24}}\tilde{q}^{\frac{1}{2}\boldsymbol{\tilde{Q}}^{2}}\tilde{y}^{%
\boldsymbol{\tilde{Q}}}\right) \right] \\
&=&Tr\left[ \left( q^{\hat{h}-\frac{c}{24}}q^{\frac{1}{2}\left( \boldsymbol{Q%
}+\rho \right) ^{2}}y^{\boldsymbol{Q}\mathbf{+}\rho }\right) \left( \tilde{q}%
^{-\widehat{\tilde{h}}-\frac{c}{24}}\tilde{q}^{\frac{1}{2}\left( \boldsymbol{%
\tilde{Q}}\mathbf{+}\tilde{\rho}\right) ^{2}}\tilde{y}^{\boldsymbol{\tilde{Q}%
}\mathbf{+}\tilde{\rho}}\right) \right]
\end{eqnarray}%
showing\textrm{\ }that, for fixed $\hat{h}$ $,$ $\widehat{\tilde{h}}$, the
spectrum is invariant under translations by charge vectors in the lattice $%
\Gamma $ namely
\begin{equation}
\begin{tabular}{lllll}
$\boldsymbol{Q}$ & $\rightarrow $ & $\boldsymbol{Q}^{\prime }$ & = & $%
\boldsymbol{Q}+\mathbf{\rho }$ \\
$\boldsymbol{\tilde{Q}}$ & $\rightarrow $ & $\boldsymbol{\tilde{Q}}^{\prime
} $ & = & $\boldsymbol{\tilde{Q}}+\mathbf{\tilde{\rho}}$%
\end{tabular}%
\quad \ \quad \text{for }\quad (\mathbf{\rho },\mathbf{\tilde{\rho}})\in
\Gamma  \label{map}
\end{equation}%
One can therefore encodes the whole spectrum of the theory via the spectral
flow \cite{pol2}$.$ The integer lattice $\Gamma ^{\ast },$ \textrm{which is}
\textrm{dual to the even integer charge lattice }$\Gamma ,$ is given by:%
\begin{equation}
\Gamma ^{\ast }=\{(\boldsymbol{Q},\boldsymbol{\tilde{Q}})\quad |\quad
\forall \left( \mathbf{\rho },\mathbf{\tilde{\rho}}\right) \in \Gamma \quad
\mathbf{\rho .Q}-\mathbf{\tilde{\rho}.\tilde{Q}}\in \mathbb{Z\}}
\end{equation}%
It is generally different from $\Gamma $\ in the sense that it contains it
as a sublattice ($\Gamma \subseteq \Gamma ^{\ast }$); the charge vectors $%
\left( \mathbf{\rho },\mathbf{\tilde{\rho}}\right) $ obey the usual even
integer constraint relation $\mathbf{\rho }^{2}-\mathbf{\tilde{\rho}}^{2}\in
2\mathbb{Z}$. To build the spectrum, we start with the (massless and
chargeless) graviton state $|0,\mathbf{0}>$ with vanishing $\boldsymbol{Q}=%
\boldsymbol{\tilde{Q}}=0$, the above map (\ref{map}) generates a tower of
quantum states with \emph{weight} and \emph{charge} related as follows \cite%
{MI, Rwgc}:%
\begin{equation}
\frac{\alpha ^{\prime }}{4}\text{\textsc{m}}^{2}=\frac{1}{2}\max (%
\boldsymbol{Q}^{2},\boldsymbol{\tilde{Q}}^{2})-1\qquad \Rightarrow \qquad
\frac{\alpha ^{\prime }}{4}\text{\textsc{m}}^{2}<\frac{1}{2}\max (%
\boldsymbol{Q}^{2},\boldsymbol{\tilde{Q}}^{2})  \label{c1}
\end{equation}%
reminiscent of the mass formula of a toroidal Narain compactification of the
bosonic string. In \cite{wgc3, MI, sub}, it was proved that these states are
super-extremal and do exist below the black hole threshold as they verify:
\begin{equation}
\frac{\alpha ^{\prime }}{4}M_{BH}^{2}\quad \geq \quad \frac{1}{2}\max (%
\boldsymbol{Q}^{2},\boldsymbol{\tilde{Q}}^{2})\quad >\quad \frac{\alpha
^{\prime }}{4}\text{\textsc{m}}^{2}  \label{C2}
\end{equation}%
This constraint represents the sublattice weak gravity conjecture in AdS$%
_{3} $ with superextremal particles populating the sublattice $\Gamma ^{\ast
}.$ It must be emphasised that the superextremal states below the black hole
threshold stem from the gauged symmetry $SL(2,\mathbb{R})\times SL(2,\mathbb{%
R})\times U(1)^{p+q-2}.$ At finite distances of the moduli space, this
symmetry is gauged and the associated states will not be included in the
ensemble averaging governed by global symmetries. And the superextremal
states $\boldsymbol{Q}\neq \boldsymbol{0}$, $\boldsymbol{\tilde{Q}}\neq
\boldsymbol{0}$ of the sublattice $\Gamma ^{\ast }$ can be identified with
the fluctuations away from the average given by the deviations:%
\begin{equation}
f\left( \tau ,\tilde{\tau};\mathbf{\mu },\mathbf{\ \tilde{\mu}}\right) =\tau
_{2}^{\left( p+q\right) /4}\left[ \left\langle \vartheta \left( m,\tau ,%
\tilde{\tau};\mathbf{\mu },\mathbf{\ \tilde{\mu}}\right) \right\rangle _{%
\mathcal{M}_{\mathfrak{Q}}}-E_{s}\left( \tau ,\tilde{\tau}\right) \right]
\neq 0  \label{fluc}
\end{equation}

If we disregard the fugacities ($\mathbf{\mu \rightarrow 0};$\textbf{\ }$%
\mathbf{\tilde{\mu}\rightarrow 0}),$ the fluctuations diminish. The states
contributing to the averaging are above the black hole threshold as they are
subextremal states that violate the WGC. There is in consequence no ensemble
averaging below the black hole threshold and only heavy states (\textsc{m }$%
\geq M_{BH}$) are averaged. Therefore, the fluctuations corresponding to
light states (\textsc{m }$\leq M_{BH}$) below the black hole threshold allow
small black holes in AdS$_{3}$ to evaporate; see Figure \textbf{\ref{lattice}%
}:
\begin{figure}[h]
\begin{center}
\includegraphics[width=10cm]{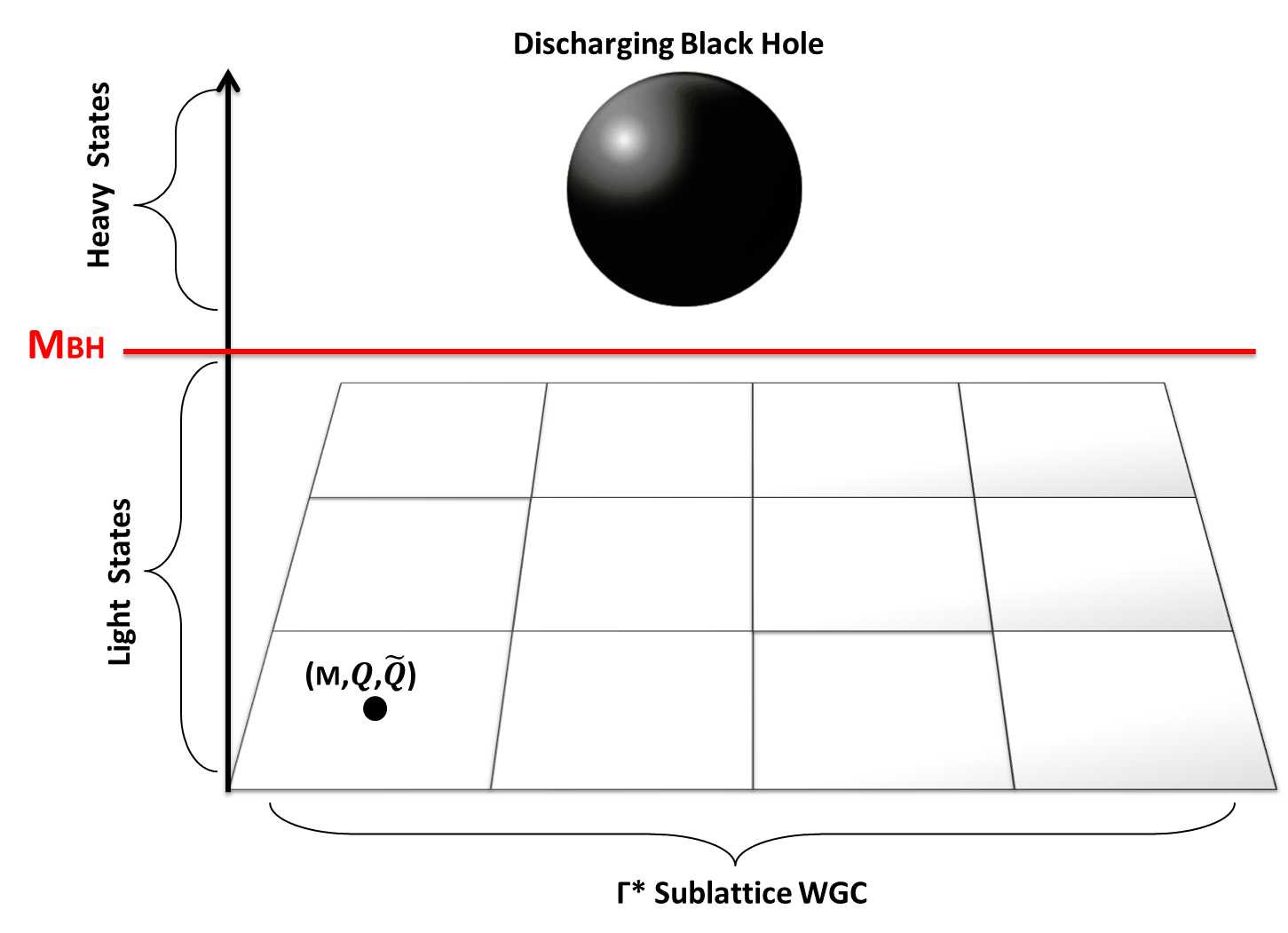}
\end{center}
\par
\vspace{-0.5cm}
\caption{Charged states $|$\textsc{m}$,\boldsymbol{q,\tilde{q}}>$ below the
black hole threshold $M_{BH}$ populates a sublattice in the dual charge
lattice $\Gamma ^{\ast }$ verifying the WGC. These states amplify the
fluctuations away from the average.}
\label{lattice}
\end{figure}

In a consistent theory of 3D quantum gravity, one must therefore consider
fluctuations besides the ensemble of averaged CFTs not only to break the
emerging global symmetries as we have seen previously, but to also generate
super-extremal states so that we can insure the evaporation of black holes.

To provide further arguments on the identification (\ref{C2})=(\ref{fluc}),
we study the torus density of states $\rho \left( \Delta ,\tilde{\Delta},%
\mathbf{\nu },\mathbf{\tilde{\nu}}\right) $ to show that light states below
the black hole threshold do in fact dominate the fluctuations away from the
averaging of the ensemble.

\subsection{Torus density of states}

In this part, we aim to study the spectrum of states at the asymptotic limit
of the 3D manifold having a torus boundary. We first consider the partition
function (\ref{ZB}) in terms of the density of states as follows%
\begin{eqnarray}
Z_{Y} &=&Tr\left( q^{L_{0}-\frac{c}{24}}y^{\boldsymbol{\hat{Q}}_{L}}q^{-\bar{%
L}_{0}-\frac{c}{24}}\tilde{y}^{\boldsymbol{\hat{Q}}_{R}}\right) \\
&=&\sum_{\Delta ,\tilde{\Delta},\mathbf{\nu },\mathbf{\tilde{\nu}}}\rho
(\Delta ,\tilde{\Delta},\mathbf{\nu },\mathbf{\tilde{\nu})}e^{2\pi i\Delta
\tau }e^{-2\pi i\tilde{\Delta}\tilde{\tau}}e^{2\pi i\mathbf{\mu }.\mathbf{%
\nu }}e^{-2\pi i\mathbf{\tilde{\mu}}.\mathbf{\tilde{\nu}}}
\end{eqnarray}%
Here, $\rho (\Delta ,\tilde{\Delta},\mathbf{\nu },\mathbf{\tilde{\nu})}$\
denotes the number of states characterised by conformal weights $\Delta ,$ $%
\tilde{\Delta}$\ and charge vectors\textbf{\ }$\mathbf{\nu },$ $\mathbf{%
\tilde{\nu}}$; they correspond to the\textrm{\ }eigenvalues of the conformal
($L_{0},\bar{L}_{0}$) and\ the charge ($\boldsymbol{\hat{Q}}_{L},\boldsymbol{%
\hat{Q}}_{R}$)\textrm{\ }operators on the eigenstates\textrm{\ }$\left\vert
\Delta ,\mathbf{\nu }\right\rangle \otimes \left\vert \tilde{\Delta},\mathbf{%
\tilde{\nu}}\right\rangle $:%
\begin{equation}
\begin{tabular}{lll}
$L_{0}\left\vert \Delta ,\tilde{\Delta},\mathbf{\nu },\mathbf{\tilde{\nu}}%
\right\rangle $ & $=$ & $\Delta \left\vert \Delta ,\tilde{\Delta},\mathbf{%
\nu },\mathbf{\tilde{\nu}}\right\rangle $ \\
$\boldsymbol{\hat{Q}}_{L}\left\vert \Delta ,\tilde{\Delta},\mathbf{\nu },%
\mathbf{\tilde{\nu}}\right\rangle $ & $=$ & $\mathbf{\nu }\left\vert \Delta ,%
\tilde{\Delta},\mathbf{\nu },\mathbf{\tilde{\nu}}\right\rangle $%
\end{tabular}%
\end{equation}%
and similarly for the right homologue. The density of states is approximated
given by the extended Cardy's formula reading as follows \cite{cardy}:%
\begin{equation}
\log \left[ \rho (\Delta ,\tilde{\Delta},\mathbf{\nu },\mathbf{\tilde{\nu})}%
\right] \approx 2\pi \sqrt{\frac{c_{L}}{6}\left( \Delta -\frac{c_{L}}{24}-%
\frac{\mathbf{\nu }^{2}}{6k}\right) }+2\pi \sqrt{\frac{c_{R}}{6}\left(
\tilde{\Delta}-\frac{c_{R}}{24}-\frac{\mathbf{\tilde{\nu}}^{2}}{6\tilde{k}}%
\right) }  \label{cardy}
\end{equation}%
where $c_{L}=p,$ $c_{R}=q$ and charged vectors ($\mathbf{\nu }$, $\mathbf{%
\tilde{\nu}}$ ) of the symmetry group $U(1)^{p-1}\times $ $U(1)^{q-1}.$

\subsubsection{Characteristic properties of $\protect\rho (\Delta ,\tilde{%
\Delta},\mathbf{\protect\nu },\mathbf{\tilde{\protect\nu})}$}

\begin{figure}[h]
\begin{center}
\includegraphics[width=10cm]{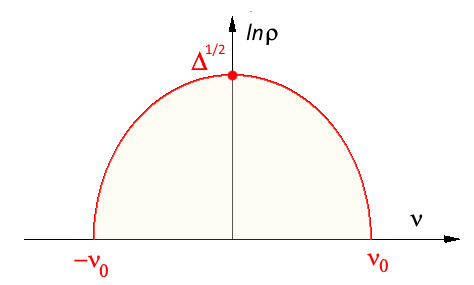}
\end{center}
\par
\vspace{-0.5cm}
\caption{A schematic representation of the allowed U(1) charges for quantum
eigenstates\textrm{\ }$\left\vert \Delta ,\mathbf{\protect\nu }\right\rangle
$ for a given weight $\Delta ^{\prime }$ and varying charge $\protect\nu $
(pink area)$.$ The boundary charge values are given by $\pm \mathbf{\protect%
\nu }_{0}=\pm \protect\sqrt{6k\Delta ^{\prime }}.$}
\label{rho}
\end{figure}

To declutter the above Cardy formula, let us set $\Delta ^{\prime }=\Delta -%
\frac{c_{L}}{24}$ and $\tilde{\Delta}^{\prime }=\tilde{\Delta}-\frac{c_{R}}{%
24}$ to remove the shifts by c/24 for a more streamlined discussion.
Substituting into (\ref{cardy}), the logarithm of the density becomes%
\begin{equation}
\log \left[ \rho (\Delta ^{\prime },\tilde{\Delta}^{\prime },\mathbf{\nu },%
\mathbf{\tilde{\nu})}\right] \approx 2\pi \sqrt{\frac{p}{6}\Delta ^{\prime
}\left( 1-\frac{\mathbf{\nu }^{2}}{6k\Delta ^{\prime }}\right) }+2\pi \sqrt{%
\frac{q}{6}\tilde{\Delta}^{\prime }\left( 1-\frac{\mathbf{\tilde{\nu}}^{2}}{6%
\tilde{k}\tilde{\Delta}^{\prime }}\right) }  \label{rx}
\end{equation}%
indicating that $\rho $ is a function of $(\Delta ^{\prime },\tilde{\Delta}%
^{\prime })$ and the charge vectors $(\mathbf{\nu },\mathbf{\tilde{\nu}}).$
This expression exhibits a remarkable constraint on the allowed values of
the density in terms of $(\Delta ^{\prime },\tilde{\Delta}^{\prime },\mathbf{%
\nu },\mathbf{\tilde{\nu}).}$ Using the positivity of $(\Delta ^{\prime },%
\tilde{\Delta}^{\prime })$ for unitary representations, and the fact that
U(1) is a compact symmetry, we conclude that $k,\tilde{k}>0.$ Therefore, not
all charge vectors $(\mathbf{\nu },\mathbf{\tilde{\nu}})$ are permitted.
This restriction arises from the minus signs in the square roots of the
above equation (\ref{rx}). For fixed weights $(\Delta ^{\prime },\tilde{%
\Delta}^{\prime }),$ the allowed charges must satisfy the conditions:
\begin{equation}
\frac{\mathbf{\nu }^{2}}{\mathbf{\nu }_{0}^{2}}\leq 1\qquad ,\qquad \frac{%
\mathbf{\tilde{\nu}}^{2}}{\mathbf{\tilde{\nu}}_{0}^{2}}\leq 1  \label{nu}
\end{equation}%
where we set $\mathbf{\nu }_{0}^{2}:=6k\Delta ^{\prime }$ and $\mathbf{%
\tilde{\nu}}_{0}^{2}:=6\tilde{k}\tilde{\Delta}^{\prime }$. The permitted
values form a compact set given by the cross product of two half-balls $%
\mathbb{B}_{L}^{p}\times \mathbb{B}_{R}^{q}$ with radii $\left\Vert \mathbf{%
\nu }_{0}\right\Vert $ and $\left\Vert \mathbf{\tilde{\nu}}_{0}\right\Vert $
respectively [recall that $\mathbf{\nu }=\left( \nu ,\cdots ,\nu _{p}\right)
$ and $\mathbf{\tilde{\nu}}=\left( \tilde{\nu},\cdots ,\tilde{\nu}%
_{q}\right) $]. A schematic representation of the half $\mathbb{B}_{L}^{p}$
is depicted in Figure \textbf{\ref{rho}}
For the special case $\Delta ^{\prime }=\tilde{\Delta}^{\prime }=0$,
corresponding to massless states, the charge vectors $\mathbf{\nu }$ and $%
\mathbf{\tilde{\nu}}$ must also vanish. In this instance, the corresponding
quantum state is given by the ground state $\left\vert 0,\mathbf{0}%
\right\rangle $\textrm{\ }describing the graviton with a density equal to one%
\textrm{\ }($\rho =1)$\textrm{.}

Furthermore, by setting $k=\tilde{k}=1$, one can rewrite the constraint
relations (\ref{nu}) like
\begin{equation}
\begin{tabular}{lll}
$\mathbf{\nu }^{2}+\mathbf{\tilde{\nu}}^{2}$ & $\leq $ & $\mathbf{\nu }%
_{0}^{2}+\mathbf{\tilde{\nu}}_{0}^{2}$ \\
$\mathbf{\nu }^{2}-\mathbf{\tilde{\nu}}^{2}$ & $\leq $ & $\mathbf{\nu }%
_{0}^{2}-\mathbf{\tilde{\nu}}_{0}^{2}$%
\end{tabular}%
\qquad \Rightarrow \qquad
\begin{tabular}{lll}
$\mathbf{\nu }^{2}+\mathbf{\tilde{\nu}}^{2}$ & $\leq $ & $6(\Delta ^{\prime
}+\tilde{\Delta}^{\prime })$ \\
$\mathbf{\nu }^{2}-\mathbf{\tilde{\nu}}^{2}$ & $\leq $ & $6(\Delta ^{\prime
}-\tilde{\Delta}^{\prime })$%
\end{tabular}%
\end{equation}%
Putting $\mathbf{\nu }^{2}=12\boldsymbol{Q}^{2}$ and $\mathbf{\tilde{\nu}}%
^{2}=12\boldsymbol{\tilde{Q}}^{2}$ as well as $\mathbf{\nu }^{2}/\mathbf{\nu
}_{0}^{2}=2\boldsymbol{Q}^{2}/\Delta ^{\prime }$ and $\mathbf{\tilde{\nu}}%
^{2}/\mathbf{\tilde{\nu}}_{0}^{2}=2\boldsymbol{\tilde{Q}}^{2}/\tilde{\Delta}%
^{\prime }\ $in the above relationships, we obtain:\
\begin{equation}
\begin{tabular}{lll}
$\boldsymbol{Q}^{2}+\boldsymbol{\tilde{Q}}^{2}$ & $\leq $ & $\frac{1}{2}%
(\Delta ^{\prime }+\tilde{\Delta}^{\prime })$ \\
$\boldsymbol{Q}^{2}-\boldsymbol{\tilde{Q}}^{2}$ & $\leq $ & $\frac{1}{2}%
(\Delta ^{\prime }-\tilde{\Delta}^{\prime })$%
\end{tabular}%
\end{equation}%
which, after substituting $\frac{\alpha ^{\prime }}{4}\text{\textsc{m}}%
=(\Delta ^{\prime }+\tilde{\Delta}^{\prime })/2$\ and $J=(\Delta ^{\prime }-%
\tilde{\Delta}^{\prime })/2,$ lead to:%
\begin{equation}
\begin{tabular}{ccc}
$\frac{\alpha ^{\prime }}{2}\text{\textsc{m}}^{2}$ & $\geq $ & $\boldsymbol{Q%
}^{2}+\boldsymbol{\tilde{Q}}^{2}$ \\
$J$ & $\geq $ & $\boldsymbol{Q}^{2}-\boldsymbol{\tilde{Q}}^{2}$%
\end{tabular}
\label{mj}
\end{equation}%
implying in turns that%
\begin{equation}
\frac{\alpha ^{\prime }}{2}\text{\textsc{m}}^{2}\geq \max (\boldsymbol{Q}%
^{2},\boldsymbol{\tilde{Q}}^{2})  \label{jm}
\end{equation}%
Comparing this result with the inequalities (\ref{c1}-\ref{C2}), in
particular $\frac{\alpha ^{\prime }}{2}$\textsc{m}$^{2}<\max (\boldsymbol{Q}%
^{2},\boldsymbol{\tilde{Q}}^{2})$, we conclude that the states $\left\vert
\Delta ^{\prime },\mathbf{\nu }\right\rangle $ within the pink phase of the
Figure \textbf{\ref{rho}} are ruled out by the WGC. This result holds as
well for generic values of $k$ and $\tilde{k}$ because the analysis for $k=%
\tilde{k}=1$ extends straightforwardly to higher values by substituting $%
\mathbf{\nu }^{2}=12k\boldsymbol{Q}^{2}$ and $\mathbf{\tilde{\nu}}^{2}=12%
\tilde{k}\boldsymbol{\tilde{Q}}^{2};$ thus leading to the relations $\mathbf{%
\nu }^{2}/\mathbf{\nu }_{0}^{2}=2\boldsymbol{Q}^{2}/\Delta ^{\prime }$ and $%
\mathbf{\tilde{\nu}}^{2}/\mathbf{\tilde{\nu}}_{0}^{2}=2\boldsymbol{\tilde{Q}}%
^{2}/\tilde{\Delta}^{\prime }$ which are identical to those descending from
the case $k=\tilde{k}=1.$

In summary, the results obtained for the density $\rho $ have been based on
the constraints (\ref{nu}) and the derived states are found to be excluded
by the WGC. This raises an important question regarding the charged states
allowed by the WGC and their derivation from the density. This issue will be
addressed in the following discussion.

\subsubsection{Beyond Cardy formula (\protect\ref{rx})}

During the evaporation of an unstable BTZ, charged states are emitted. This
naturally raises the question: Where do these evaporated states go? This
inquiry becomes particularly relevant given that the states identified
earlier, represented by the pink region of the Figure \textbf{\ref{rho}, }%
are forbidden by the WGC.

To address this, we start by recalling that for unitary representations
characterised by positive weights ($\Delta ^{\prime },\tilde{\Delta}^{\prime
})$ and positive CS integers $k,\tilde{k},$ the U(1) charges of the
additional Chern-Simons potentials are constrained as shown in (\ref{mj}-\ref%
{jm}). Accordingly, the evaporated charged states must extend beyond these
conditions. Although not explicitly derived from Cardy's formula, we
demonstrate below that the evaporated states allowed by the WGC are indeed
encapsulated within the formula (\ref{rx}).

To derive these states, we draw inspiration from the study of phase
transitions in topological matter \textrm{\cite{TPM}-\cite{TPM3}}. Adapting
this idea to topological gravity coupled to abelian CS potentials, we
reinterpret the upper bounds on the charges in (\ref{nu}),\textrm{\ }%
\begin{equation}
\frac{\mathbf{\nu }^{2}}{\mathbf{\nu }_{0}^{2}}=1\qquad ,\qquad \frac{%
\mathbf{\tilde{\nu}}^{2}}{\mathbf{\tilde{\nu}}_{0}^{2}}=1,  \label{pb}
\end{equation}%
\textrm{\ }as permeable barriers that can be crossed by topological states.
This novel interpretation can be implemented by replacing the standard
relationship (\ref{rx}) by the following relaxed one\textrm{\ \ }%
\begin{equation}
\log \left[ \rho (\Delta ,\tilde{\Delta},\mathbf{\nu },\mathbf{\tilde{\nu})}%
\right] \approx 2\pi \sqrt{\frac{p}{6}\Delta ^{\prime }\left\vert 1-\frac{%
\mathbf{\nu }^{2}}{\mathbf{\nu }_{0}^{2}}\right\vert }+2\pi \sqrt{\frac{q}{6}%
\tilde{\Delta}^{\prime }\left\vert 1-\frac{\mathbf{\tilde{\nu}}^{2}}{\mathbf{%
\tilde{\nu}}_{0}^{2}}\right\vert }  \label{R}
\end{equation}%
where the argument $1-\mathbf{\nu }^{2}/\mathbf{\nu }_{0}^{2}$ is
substituted by its absolute value $\left\vert 1-\mathbf{\nu }^{2}/\mathbf{%
\nu }_{0}^{2}\right\vert ,$ and similarly for the right homologue. As a
result, the quantity $1-\mathbf{\nu }^{2}/\mathbf{\nu }_{0}^{2}$ gets
rotated by an angle $\varphi =\pi $ $\func{mod}2\pi $ when crossing the
barriers (\ref{pb}), causing the density $\log \rho $ to rotate in turn by
an angle $\varphi /2=\pi /2$ $\func{mod}\pi .$

\begin{figure}[h]
\begin{center}
\includegraphics[width=10cm]{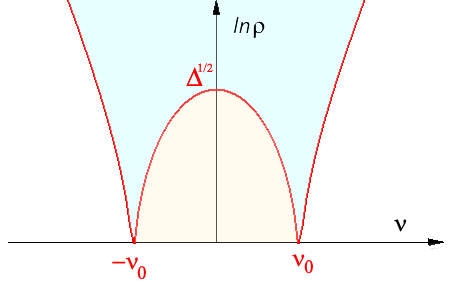}
\end{center}
\par
\vspace{-1cm}
\caption{A schematic representation of the allowed U(1) charges for the
quantum eigenstates $\left\vert \Delta ,\mathbf{\protect\nu }\right\rangle $
for permeable barrier given by eq(\protect\ref{pb}). The green area gives
the charged states allowed by the WGC.}
\label{TP}
\end{figure}

This new relaxed expression allows the charge vectors $(\mathbf{\nu },%
\mathbf{\tilde{\nu})}$ to take values beyond the bounds (\ref{nu}). In
addition to $\mathbf{\nu }^{2}<\mathbf{\nu }_{0}^{2}$ and $\mathbf{\tilde{\nu%
}}^{2}<\mathbf{\tilde{\nu}}_{0}^{2},$ we now also accommodate $\mathbf{\nu }%
_{0}^{2}<\mathbf{\nu }^{2}$ and $\mathbf{\tilde{\nu}}_{0}^{2}<\mathbf{\tilde{%
\nu}}^{2}$ corresponding to $\frac{\alpha ^{\prime }}{2}$\textsc{m}$^{2}\leq
\mathbf{\nu }^{2}+\mathbf{\tilde{\nu}}^{2}$ and further to the condition $%
\frac{\alpha ^{\prime }}{2}$\textsc{m}$^{2}\leq \max \left( \mathbf{\nu }%
^{2},\mathbf{\tilde{\nu}}^{2}\right) ,$ which can be expressed as follows
\begin{equation}
\frac{\alpha ^{\prime }}{4}\text{\textsc{m}}^{2}\leq \frac{1}{2}\max (%
\boldsymbol{Q}^{2},\boldsymbol{\tilde{Q}}^{2})
\end{equation}%
in agreement with the WGC. The set of charged states allowed by the WGC is
illustrated in Figure \textbf{\ref{TP}}, it is represented by the green
shaded area.

\textbf{Heavy and light charged states}

From the above results, we identify two distinct\textrm{\ }phases according
to the value of the weight in terms of the charges:

\textbf{A) Phase with heavy states}\newline
These states satisfy the constraints $\mathbf{\nu }^{2}\leq \mathbf{\nu }%
_{0}^{2}$ and $\mathbf{\tilde{\nu}}^{2}\leq \mathbf{\tilde{\nu}}_{0}^{2}$,
implying that they have large weights seeing that $\mathbf{\nu }^{2}\leq
6k\Delta ^{\prime }$ and $\mathbf{\tilde{\nu}}^{2}\leq 6\tilde{k}\tilde{%
\Delta}^{\prime }$. As their mass exceeds their charge, they violate the
weak gravity condition (\ref{C2}). Accordingly, they do not belong to the
charge \emph{sublattice} consistent with the WGC$.$

\textbf{B) Phase with light states}\newline
Here, the states have smaller weights sitting in the domain $\mathbf{\nu }%
^{2}\gtrsim 6k\Delta ^{\prime }$ and $\mathbf{\tilde{\nu}}^{2}\gtrsim 6%
\tilde{k}\tilde{\Delta}^{\prime }.$ They are super-extremal states that do
belong to the charge sublattice verifying the WGC condition. They describe
the emitted particles during the discharge of the charged BTZ black hole.

\begin{figure}[h]
\begin{center}
\includegraphics[width=12cm]{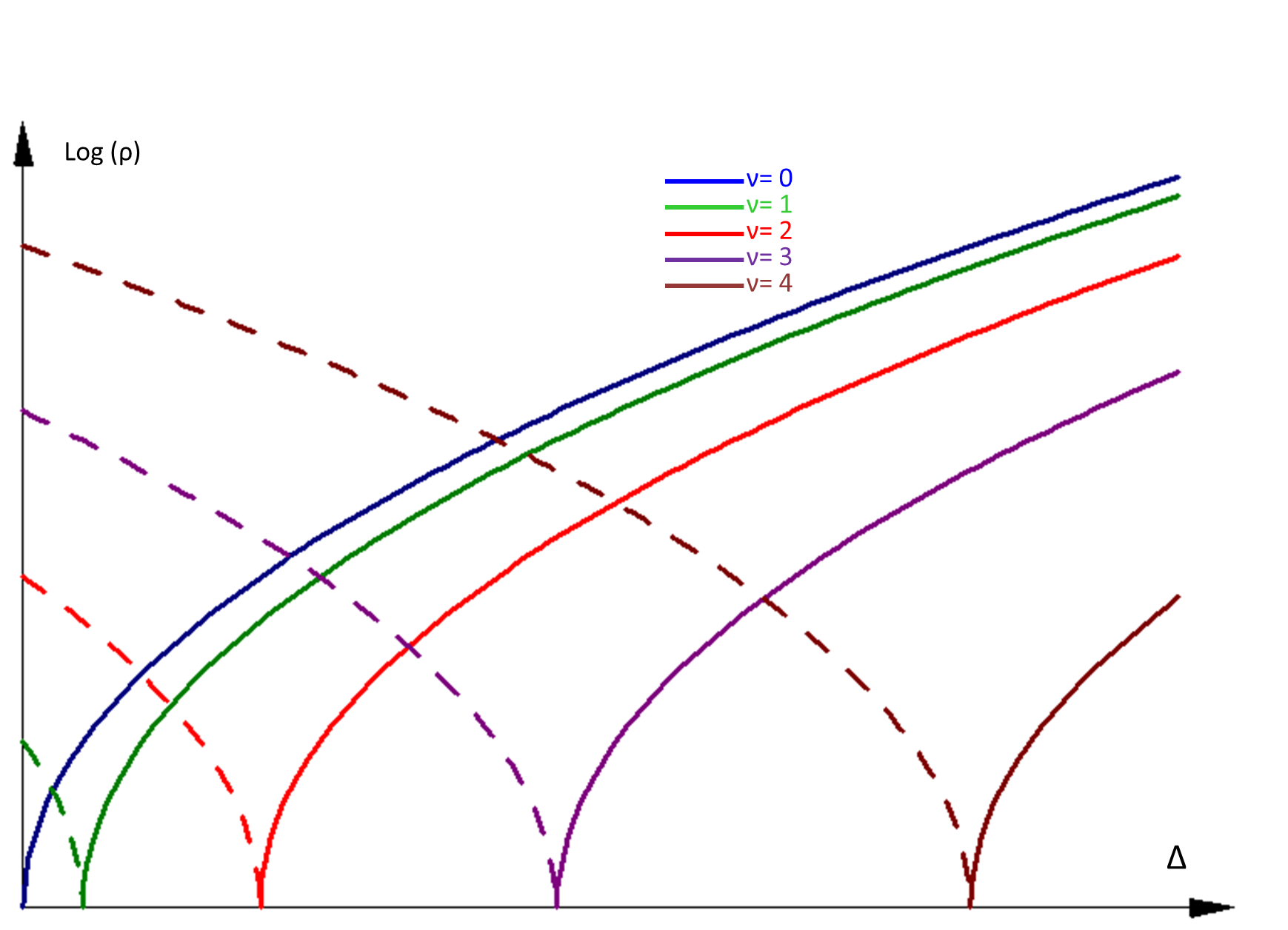}
\end{center}
\caption{Graphic representations of the logarithmic density of states in
terms of the weight $\Delta $ for different values of the charges $\mathbf{%
\protect\nu }$. The connected lines represent the function $\log \protect%
\rho =\protect\sqrt{\Delta -\frac{\mathbf{\protect\nu }^{2}}{6}}$ (heavy
states) and the dotted curves correspond to $\log \protect\rho =\protect%
\sqrt{\frac{\mathbf{\protect\nu }^{2}}{6}-\Delta }$ (light states).}
\label{ploot}
\end{figure}

To display these phases, we plot the logarithm of density of states as a
function of the weights for fixed values of the U(1) charges, focusing on
the left sector with $p=6,$ $k=1$; see Figure \textbf{\ref{ploot}}
\newpage
\textrm{For smaller weights }$\Delta $\textrm{, the growth rate of the
logarithmic density log}$\left( \rho \right) $\textrm{\ decreases as the
charges} $\mathbf{\nu }$ \textrm{increase.} \textrm{The density is therefore
highly sensitive to the charges in the light regime leading to noticeable
fluctuations.} \textrm{Considering an ensemble of }$\mathcal{GNCFT}$'s\textrm{%
, where each theory is associated with a partition function expressed in
terms of the density of states, we conclude the following:}

\begin{description}
\item[$(\mathbf{i})$] \textrm{\ The average is dominated by heavy states.}

\item[$(\mathbf{ii})$] \textrm{\ Light states, being fewer in number,
contribute less significantly in magnitude to the ensemble. However, their
contributions are more variable, leading to deviations from the average.}
\end{description}

\section{Conclusion and comments}
\label{sec5}
In conclusion, this paper builds on the results of \cite{M2} to further
explore the connection between ensemble averaging of generalised Narain CFTs
and the Swampland program. We began by performing a comparative study
between standard and generalised Narain CFTs where we highlighted the main
differences in table (\ref{tab}) based on many aspects including type of
lattices and modularity of the partition function.

Next, we studied the emergence of global symmetries from ensemble averaging
with regard to the Swampland program. As this kind of symmetries are
prohibited by the no global symmetry Swampland conjecture, one had to
consider fluctuations away from the average to circumvent them. These
fluctuations were expressed via deviations of the Siegel-Weil formula as
showed in (\ref{Fluc}). However, one must also determine the bulk origin of
such fluctuations and how they may arise from the bulk to the boundary.

In order to investigate the physical significance of the fluctuations
introduced in \cite{M2} from the perspective of the bulk physics and
identify the states giving rise to them, we first constructed the AdS$_{3}$
gravitational dual. We considered Einstein AdS$_{3}$ gravity coupled to a\
set of U(1)$^{p+q-2}$ gauge fields in the Chern-Simons formulation and
showed that upon imposing diagonal boundary conditions, the asymptotic
symmetries are given by U(1)$^{p+q}$ affine algebra (\ref{asy}) conforming
to the bulk of the averaged ensemble of generalised Narain CFTs (\ref{CS1}).

Moreover, it turned out that the new bulk theory bears charged BTZ black
hole solutions (\ref{bh}) and that an evaporating BTZ black hole emits
super-extremal states below the black hole threshold. In fact, using modular
invariance, one can prove that the super-extremal states populate a
sublattice as predicted by the Swampland sublattice weak gravity conjecture.
Furthermore, the super-extremal particles emitted from the evaporating
charged BTZ black hole generate the fluctuations around the average as
ensemble averaging only concerns heavy states that exceed the black hole
threshold. Fluctuations around this average can be therefore identified in
the bulk with the emitted light states below the black hole threshold
populating the sublattice (\ref{c1}).

As a consistency check, we studied the density of states to derive
constraints on the charged vectors. To ensure alignment with the sublattice
WGC, we proposed a revised logarithmic density based on treating upper
bounds as permeable barriers. This allowed the computation of the WGC
conditions from the density function and categorized states into two
regimes. We distinguished between heavy states, which contribute to the
average due to their large concentration, and light states that generate
fluctuations around the average.

\textrm{In the end}, it would be interesting to examine the theory before
and after the averaging. Studying the Swampland constraints on the
gravitational dual prior to averaging, in the framework of the
Maxwell-Chern-Simons theory, and then re-deriving them for the averaged
theory could offer further insights into how averaging influences the
Swampland criteria.
\begin{equation*}
\end{equation*}

\end{document}